\documentclass[fleqn,usenatbib]{mnras}

\usepackage{newtxtext,newtxmath}

\usepackage[T1]{fontenc}

\DeclareRobustCommand{\VAN}[3]{#2}
\let\VANthebibliography\thebibliography
\def\thebibliography{\DeclareRobustCommand{\VAN}[3]{##3}\VANthebibliography}
\usepackage{graphicx}	% Including figure files
\usepackage{amsmath}	% Advanced maths commands
\usepackage{xfrac,nicefrac}

\newcommand{\Mdot}{\mbox{\,$\rm M_{\odot}$}}        % solar mass
\newcommand{\Zdot}{\mbox{\,$\rm Z_{\odot}$}}        % solar metallicity
\newcommand{\ZLMC}{\mbox{\,$\rm Z_{\rm LMC}$}}        % LMC metallicity
\newcommand{\kms}{\,km s$^{-1}$}   			% kms-1
\newcommand{\aov}{$\alpha_{\rm ov}$}		% overshoot alpha
\newcommand{\amlt}{$\alpha_{\rm mlt}$}		% mlt alpha
\newcommand{\vsini}{$\varv \sin i$}

\title[Age in M-L plane]{Stellar age determination in the Mass-Luminosity Plane}
\author[E. R. Higgins et al.]{
Erin R. Higgins$^{1}$\thanks{E-mail: erin.higgins@armagh.ac.uk}, \&
Jorick S. Vink$^{1}$\\
$^{1}$Armagh Observatory and Planetarium, College Hill, Armagh BT61 9DG, N. Ireland}

\date{Accepted 14 October. Received 15 August}
\pubyear{2022}

\begin{document}

\maketitle
\begin{abstract}
\noindent
The ages of stars have historically relied on isochrone fitting of standardised grids of models. While these stellar models have provided key constraints on observational samples of massive stars, they inherit many systematic uncertainties, mainly in the internal mixing mechanisms applied throughout the grid, fundamentally undermining the isochrone method. In this work, we utilise the M-L plane of Higgins \& Vink as a method of determining stellar age, with mixing-corrected models applying a calibrated core overshooting \aov\ and rotation rate to fit the observational data. We provide multiple test-beds to showcase our new method, while also providing comparisons to the commonly-used isochrone method, highlighting the dominant systematic errors. We reproduce the evolution of individual O stars, and analyse the wider sample of O and B supergiants from the VLT-FLAMES Tarantula Survey, providing dedicated models with estimates for \aov, $\Omega/\Omega_{\rm{crit}}$, and ultimately stellar ages. The M-L plane highlights a large discrepancy in the spectroscopic masses of the O supergiant sample. Furthermore the M-L plane also demonstrates that the evolutionary masses of the B supergiant sample are inappropriate. Finally, we utilise detached eclipsing binaries, VFTS 642 and VFTS 500, and present their ages resulting from their precise dynamical masses, offering an opportunity to constrain their interior mixing. For the near-TAMS system, VFTS 500, we find that both components require a large amount of core overshooting (\aov\ $\simeq$ 0.5), implying an extended main-sequence width. We hence infer that the vast majority of B supergiants are still burning hydrogen in their cores.
\end{abstract}

% dated ...to our ...yet it is indirectly predicted from stellar parameters based on theoretical work which evolutionary stages and life

\section{Introduction}

Stellar ages are key for our understanding of galaxy evolution, chemical enrichment and yields from stellar winds, as well as population synthesis studies. The ages of individual stars can not be measured directly, however, and rely on fitting to theoretical models. The most common method of ageing clusters of stars has historically been to utilise Colour-Magnitude diagrams \citep[e.g.][]{Massey95}, motivated by observational measurements which do not require spectroscopic analysis. Individual stars with spectral classes and luminosities calculated from stellar atmosphere modelling, however, can provide direct comparison to stellar evolution models to provide discrete stellar ages \citep{Soder10}. 

The commonly-used Hertzsprung-Russell diagram (HRD) has widely been accepted as the method of comparing observations and theoretical models, while showcasing the inconsistencies and similarities between the observational data and model grids \citep[e.g.][]{WB97,Wright10}. Isochrones are then used to fit the model grid to the observational data providing a constraint on the age \citep{Nord04}. While the HRD provides useful information about stellar observables such as luminosity and surface temperature, it does not provide a comparison of theoretical processes such as internal mixing or the effects of stellar winds. In fact, the standardised physics of model grids do not take into account that a population of stars will have different rotation rates and convective core sizes, which will directly affect the path that this stellar population will take and how long stars will spend in various regions of the HRD. This suggests that the fundamental method of dating stars by fitting single isochrones to observations is flawed, since stellar evolution models are subject to a range of inputs, each with their inherent uncertainties, including convective core overshooting and rotational mixing \citep[e.g.][]{georgy14}. The drawbacks of the isochrone method indicate that a new method which incorporates constraints from both observational measurements and theoretical inputs is required for accurate stellar ageing. In \cite{Higgins} an alternative tool which disentangles the effects of mixing and mass loss in models by comparing the mass and luminosity was developed as a novel method of comparing models with observations. In this study we further develop the Mass-Luminosity (M-L) Plane, now to estimate the age of stars. 

The theory of massive star evolution has been progressing over the last decade with the aid of large observational samples of massive stars, providing a robust comparison to grids of stellar evolution models. Such observations have also provided an opportunity to compare different grids of stellar models calculated with varying codes and default implementations of physical processes \citep[e.g.][]{Bonn11, Gen12, Chieffi, MIST}. Studies have showcased systematic differences in codes which are highlighted by increased luminosities and masses in the \texttt{GENEC} model grids \citep{Gen12} compared to the grid of \texttt{STERN} models \citep{Bonn11}. Moreover, model atmosphere codes such as \texttt{FASTWIND} and \texttt{TLUSTY} have been tested in their spectroscopic analysis of the VLT-FLAMES Tarantula Survey (VFTS) sample, providing a comparison of the spectroscopic masses determined by various spectral models \citep{McEvoy}. These various code comparisons have led to a well-known inconsistency between estimates of the stellar mass from evolutionary models and spectroscopic models. This discrepancy between spectroscopic masses and evolutionary masses is now known as the ``mass discrepancy problem'' \citep{Herrero}. Since the masses of $M_{*}$ $>$ 20\Mdot\ stars are increasingly uncertain, the use of detached binaries provides an important tool in better constraining the mass and evolution of such massive stars. In fact, since dynamical masses from detached eclipsing binaries provide an accurate measurement of masses from system dynamics \citep{torres10, Tka}, this method may be key in solving the mass discrepancy problem.

Previous studies of detached eclipsing binaries, in the mass range of $\sim$2-15\Mdot, have provided estimates of the core extension by convective overshooting \citep{bressan12, Stancliffe, ClaretTorres, ConstantinoBaraffe}, finding an increase in \aov\ with increased mass, with \cite{ClaretTorres18} finding a plateau at $\sim$ 2\Mdot. Moreover, \cite{Costa19a} provide analysis of detached eclipsing binaries with a Baysian approach, in which they calibrate the interior mixing due to rotation and overshooting, finding a wide dispersion of mixing. The dispersion was found to be either due to a variation in \aov\ $=$ 0.3-0.8, or a constant \aov\ with a range of initial rotation rates in which the latter is proposed as the best fit. The method presented in \cite{Costa19b} provides another solution to compare stellar observations with non-standardised grids of models, where a larger proportion of the population may be represented.
% The largest homogeneous sample of massive star observations was obtained by the VLT-FLAMES Tarantula Survey (VFTS) \citep{Evans11} providing multi-epoch spectroscopy of over 800 massive stars in the Tarantula Nebula of the Large Magellanic Cloud. While other surveys have been performed on massive stars, these are largely photometric or star formation studies and as such the VFTS sample provides a much needed spectroscopic survey of massive stars. The VFTS ESO Large Programme was motivated by the lack of massive star observations and the need to better understand their binarity, evolutionary status and physical processes (such as mass loss and internal mixing). With variations in radial velocities, many massive binary systems were detected. \cite{Sana13} found that the binarity of massive stars in the VFTS sample was $\sim$ 50\% suggesting that multiplicity is an important factor in massive star evolution. Though \cite{Sana12} found $\sim$ 70\% of massive stars were in binary or multiple systems, this sample was much smaller with only 20\% of the \cite{Sana13} sample size. \cite{Taylor} suggested that non-interacting binaries would provide a valuable insight into this evolution, mainly due to reliable estimates of the mass of each component. Moreover, the age and evolutionary tracers including chemical enrichment and mass loss history may be better understood. In \cite{Higgins}, we utilised this technique by reproducing the evolution of the galactic detached binary HD166734 with constraints on the internal mixing processes and mass-loss rates.

In this study we present a new method of calculating the age of individual stars in the M-L plane, with comparison to isochronal ages. We compare evolutionary and spectroscopic masses of O and B supergiants from the VFTS, determining discrepancies in the M-L plane. We reproduce the evolution of 2 detached eclipsing binary systems in the VFTS sample, VFTS 500 and 642. We present our methods in Sect. \ref{method}, with stellar modelling outlined in Sect. \ref{models}, an introduction of the M-L plane in Sect. \ref{MLmethod} and observations from the VFTS sample in Sect. \ref{datasect}. We highlight our results in Sect. \ref{results}, with our new method of estimating stellar age in Sect. \ref{agesect}, the mass discrepancy of O and B supergiants in Sect. \ref{lmcvfts}, and the evolution of detached systems in Sect \ref{binaries}. We then provide our conclusions in Sect. \ref{conclusions}.

\section{Method}\label{method}

\subsection{Stellar models}\label{models}
In this study we avail of the public access stellar evolution code Modules for Experiments in Stellar Astrophysics \citep[\texttt{MESA}, v8845,][]{Pax15}. We have previously calculated a grid of Galactic stellar evolution models in \cite{Higgins}, and have subsequently calculated a comparable grid of models in this work corresponding to 50\% \Zdot\ for the LMC. We have calculated models with initial masses of 8\Mdot, 12\Mdot, 16\Mdot, 20\Mdot, 25\Mdot, 30\Mdot, 35\Mdot, 40\Mdot, 45\Mdot, 50\Mdot, 55\Mdot\ and 60\Mdot. We calculate comparable models in addition to these masses in order to reproduce the detached eclipsing binaries.

Convection is employed via the mixing length theory (MLT) developed by \cite{CG68} where \amlt $=$ 1.5, while implementing the Ledoux criterion denoted by $\nabla_{\rm{rad}}$  $<$ $\nabla_{\rm{ad}}$  $+$ $\frac{\phi}{\delta}$ $\nabla_{\mu}$, but where $\nabla_{\mu}$ $=$ 0, the Schwarzschild criterion is effective. We investigate the efficiency of convective core overshooting by including the step overshooting implementation in \texttt{MESA}. This method extends the core by a fraction \aov\ of the pressure scale height $H_{\mathrm{p}}$. The effects of convective core overshooting are tested for core H-burning with variations of \aov\ $=$ 0.1 and 0.5. We adopt a scaled-solar metallicity for the LMC with $Z$ $=$ 0.0088, though other studies have utilised an alternative initial $[{N/H}]$ abundance \citep[e.g.][]{Bonn11}.

The \cite{Vink01} mass loss recipe is implemented for mass loss during the hot, hydrogen-rich phases of evolution (i.e. $T_{\rm eff}$ $>$ 10kK, $X_{\rm s}$ $>$ 0.7) with a scaling factor of unity as concluded by \cite{Higgins} in our previous work, and \cite{deJager} implemented for cool stars ($T_{\rm eff}$ $<$ 10kK). The effects of rotationally-induced mass loss are discarded \citep{MullerVink, Higgins}. Rotation is included for initial rates of $\Omega$/$\Omega_{\rm crit}$ $=$ 0.1 and 0.4, with rotational instabilities employed for angular momentum transfer and chemical mixing as described by \cite{Heger00}. 

Models have been calculated with typical resolution parameters as outlined in the MESA instrument documentation \citep{Pax13}, in particular with a temporal resolution of \texttt{varcontroltarget} \texttt{= 1d-4} and spatial resolution \texttt{meshdelta} \texttt{= 1.5}. The effects of superadiabaticity via the \texttt{MESA} parameter \texttt{MLT$++$} have been omitted for all models.

\subsection{Mass - Luminosity Plane}\label{MLmethod}
In \cite{Higgins} a unique method of calibrating massive star evolution was developed, by comparing the stellar masses with observed luminosities we can disentangle stellar winds from internal mixing processes, finding upper limits to $\dot{M}$, \vsini\ and \aov. We calibrated the evolution of a detached Galactic test-bed binary, HD166734 \cite{Mahy2017}, using the M-L plane with surface nitrogen abundances as a function of observed $M_{\rm spec}$, $L$, $T_{\rm eff}$, and \vsini. The M-L plane is a particularly useful tool for analysing theoretical models and providing constraints on observational samples, ultimately improving our understanding of the dominant effects on the evolution of massive stars. Disentangling processes such as mass loss and mixing (rotation or convection) is key for improving stellar modelling since many of these dominant processes remain uncertain and are usually implemented with default settings for an entire model grid. Yet, studies from asteroseismology and spectroscopic analyses suggest that there may be a mass or metallicity dependence on internal mixing processes \citep[e.g.][]{Castro14, bowman20, scott21}.

\begin{figure}
\centering
\includegraphics[width = \columnwidth]{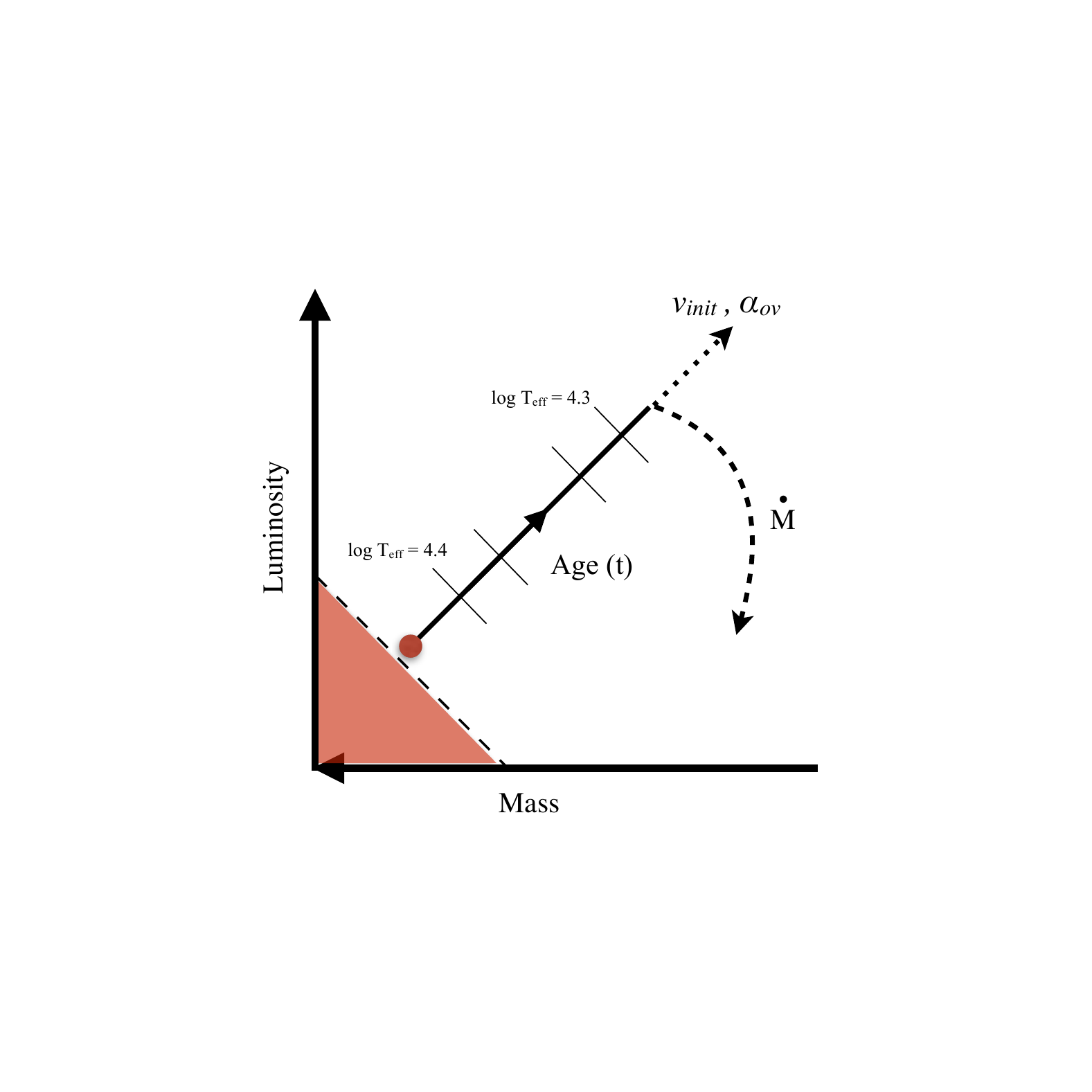}
\caption[Age in the M-L plane]{\footnotesize The Mass-Luminosity plane illustrated with an evolutionary model extending to higher luminosities and lower masses with age until an observed effective temperature is reached. Effective temperatures are denoted by markers along the evolutionary track, highlighting evolution towards cooler effective temperatures with increased evolutionary time. The red forbidden zone marks the initial mass and luminosity of each model based on the mass-luminosity relation.}
\label{age}
\end{figure}

Figure \ref{age} shows that just as in a HRD, an evolutionary track evolves left to right along the vector with decreasing effective temperature or increasing time. The mass-luminosity relation provides a forbidden region which sets the initial starting point in the M-L plane whereby the model evolves from the Zero-Age-Main-Sequence (ZAMS) through the MS towards the Terminal-Age-Main-Sequence (TAMS). The vector length in the M-L plane can be extended with extra mixing in order to reach the same effective temperature, i.e. the length of a 20\Mdot\ model at log $T_{\rm{eff}}$ $=$ 4.4 is shorter for \vsini\ $=$ 100\kms\ than for the same model with \vsini\ $=$ 300\kms. On the other hand, the gradient of the vector is solely reliant on the mass-loss rate, so a steeper gradient corresponds to a lower mass-loss rate, and a shallow gradient corresponds to a higher mass-loss rate. This effect can also be seen with increased metallicity ($Z$) due to the effects of Z-dependent winds, such that a 20\Mdot\ model calculated at \Zdot\ will have a shallow gradient in the M-L plane when compared to the same model calculated for \ZLMC\ since winds are reduced at lower $Z$. In this work we further develop the M-L plane for determining the age of individual stars, detached binary systems and large stellar populations.

\subsection{Observations}\label{datasect}
A multi-epoch survey of over 800 massive stars was completed by the VFTS \citep{Evans11}, an ESO Large Programme which has provided comprehensive information on O- and B-type stars in the 30 Doradus (30Dor) region of the Tarantula Nebula of the LMC. A key motivation of the survey was to sample the binary fraction and systems within the 30Dor region, now studied within the Tarantula Massive Binary Monitoring programme \citep[TMBM][]{TMBM1, TMBM3}. Alongside spectral properties of massive stars, the VFTS programme also investigated the rotational properties of the sample, with key insights from chemical enrichment \citep{Grin17} and the distribution of \vsini\ \citep{Ramirez13}. 

\subsubsection{O and B supergiants}
In this work, we investigate the stellar properties of O supergiants from \cite{Ramirez13} and B supergiants from \cite{McEvoy}. We also study the evolution of detached eclipsing binaries from the TMBM sample \cite{TMBM3}. Homogeneous observational samples such as the VFTS programme allow for large-scale comparison of spectral modelling and evolutionary modelling, providing an opportunity for advancing our implementation of physical processes in theoretical models. We consider single stars only in these stellar populations in order to directly probe the effects of mixing and mass loss without contamination from interacting binary effects. However, we separately compare the evolution of detached eclipsing binaries as a proxy for single star evolution prior to interaction.

\cite{Grin17} provides analysis of 72 O-type giants and supergiants, with evolutionary masses of 10-94\Mdot. These O stars are MS, H-burning stars, which produce Nitrogen ($N$) in their cores via the CNO-cycle, and display N-enhancement at their surfaces later in their MS evolution, either due to rotational mixing dredging $N$ to the surface, or due to stellar winds stripping their envelope and exposing fusion products at the stellar surface. A combined effect of both internal mixing and external winds is likely causing the surface enrichment of massive stars. Since \cite{Grin17} highlights uncertainties in the spectroscopic masses of the O supergiant sample, they utilise the evolutionary masses estimated from the \cite{Bonn11} tracks interpolated with the \texttt{BONNSAI} tool. In order to compare the spectroscopic and evolutionary masses, we include the data analysis of the O supergiants by \cite{Ramirez13} in our comparisons with the B supergiant sample.

\cite{McEvoy} details model atmosphere calculations of 34 single B supergiants with the non-local thermodynamic equilibrium (NLTE) code, \texttt{TLUSTY}, providing surface $N$ abundances, rotational velocities, and stellar parameters such as effective temperatures, spectroscopic masses, evolutionary masses and luminosities. Analysis of the B supergiant sample by \cite{McEvoy} suggested evidence for an extended MS towards an effective temperature of $\sim$22,000K. Evolved B supergiant stars are expected to lie close to the Terminal-age Main-Sequence (TAMS), but questions have been raised over the last few decades about their core burning stage due to challenges in determining the MS-width. While mainly considered post-MS, core He-burning objects due to their lowered \vsini\ estimates, this may be an effect of bi-stability braking, outlined in \cite{Vink10} as a result of increased mass-loss rates lowering the angular momentum of objects crossing the bi-stability jump around 22,000K.

\begin{table}
    \centering
    \begin{tabular}{|c|c|c|c|c|c|}
%     # Teff	log g	log L	N/H	Mevol	Mspec	Age	/Mupper	/Mlower	vsini	\C/H	\O/H	\Nerr	\Cerr	\Oerr	\Lerr	\t4g		logteff		\t4gerr	\logtefferr	star
% 44540	4.20	5.44	6.892	40.40	44.6	0.8	5.4	4.4	221.5	8.079	8.431	0.705	0.477	0.732	0.092	14.8705999	4.648750213	0.04 	0.0135		063P
% 39260	4.18	4.94	6.924	25.40	22.41	1.1	4.6	3.7	95.2	8.301	8.663	0.556	0.824	1.186	0.104	14.8371996	4.593950295	0.04	0.0135		063S
% #31810	3.98	4.63	6.903	16.60	16.22	6.0	5.4	4.4	175.9	8.079	8.431	0.705	0.477	0.732	0.092	14.8705999	4.502563669	0.04 	0.0135		116P
% #28020	3.98	4.42	7.041	13.20	16.17	8.9	4.6	3.7	117.3	8.301	8.663	0.556	0.824	1.186	0.104	14.8371996	4.447468131	0.04	0.0135		116S
% #34000	3.27	6.05	8.613	68.60	63.15	2.2	5.4	4.4	56.8	8.079	8.431	0.705	0.477	0.732	0.092	14.8705999	4.531478917	0.04 	0.0135		527P
% #34700	3.35	6.00	8.544	54.20	63.09	2.6	4.6	3.7	77.2	8.301	8.663	0.556	0.824	1.186	0.104	14.8371996	4.540329475	0.04	0.0135		527S
\hline
       VFTS & $\mathrm{log{_{10}}} \, T_{\mathrm{eff}}$  & $\mathrm{log{_{10}}}\;(L/\rm{L}_\odot)$&  $M_{\mathrm{dyn}}/M_{\odot}$	& $t_{\rm{ML}}$ (\rm{Myr})\\
       \hline \hline
       500P  & 4.61 & 5.29 & 25.1 & 6.37\\
       500S  & 4.59 & 5.21 & 23.8 & 6.41\\
       642P  & 4.61 & 5.08 & 29.8 & 2.32\\
       642S  & 4.54 & 4.68 & 19.2 & 2.28\\
        \hline
       
    \end{tabular}
    \caption{Stellar parameters of detached eclipsing binaries, adopted from \citet{TMBM3}, where P is the primary component and S represents the secondary component. The $t_{\rm{ML}}$ is the estimated current age and has been calculated using the M-L plane.}
    \label{tab:binaries}
\end{table}

\subsubsection{Detached binary systems}

In this study we provide new analysis of 2 detached binary systems from the TMBM sample: VFTS 500 and VFTS 642, with one system near the ZAMS and one system near the TAMS. The TMBM project has performed analysis of 82 massive binary systems in the LMC, 51 single lined and 31 double lined spectroscopic binaries \citep{TMBM1}. Spectral disentanglement was undertaken by \cite{TMBM3} in order to provide estimates of stellar parameters which we utilise in this work to better constrain their evolutionary status and study internal mixing processes (e.g. \aov) at multiple mass ranges. 
%The most massive O-star binary system VFTS 527 (R139), was revealed in the VLT-FLAMES Tarantula Survey (VFTS) \citep{Evans11}. It was discovered to be a massive spectroscopic binary system by \citet{Taylor}, with similar spectral types for the primary and secondary, classified as an O6.5Iafc supergiant and O6Iaf supergiant respectively. Further analysed as part of the Tarantula Massive Binary Monitoring (TMBM), the system was found to have slightly lower mass estimates \citep{TMBM1}, though required a disentangled spectra for a full analysis. 

\citet{TMBM3} have studied VFTS 500 and 642 in more detail providing spectroscopic and evolutionary masses, effective temperatures and surface abundances. Utilising these stellar parameters to probe stellar evolution at this mass range would provide a much better understanding of physical processes in this regime, as well as exploring the location of the TAMS for these masses. \citet{TMBM3} critically provide dynamical masses of each component of the detached eclipsing binary systems, giving accurate constraints on their evolution. Furthermore, this may provide a better understanding of main sequence evolution for 2 mass ranges at 2 metallicities, following \cite{Higgins}. We provide stellar parameters of VFTS 500 and 642 in Table \ref{tab:binaries}, adapted from \cite{TMBM3}.

\section{Results}\label{results}

In this work we present a new method of calculating the age of stars alongside grids of evolutionary models. We provide a comparison to observations of the VFTS O and B supergiant sample, and detached eclipsing binary systems. Subsequent to \cite{Higgins}, models were calculated for the range of initial masses 8-60\Mdot\ from pre-MS to core collapse unless convergence problems occurred in final evolutionary phases. We apply overshooting \aov\ of 0.1 and 0.5 to explore the location of the TAMS for O and B supergiants. Finally, we implement 2 rotation rates of $\Omega$/$\Omega_{\rm crit}$ $=$ 0.1 and 0.4 in order to best represent our sample of observations.
\begin{figure}
    \includegraphics[width = \columnwidth]{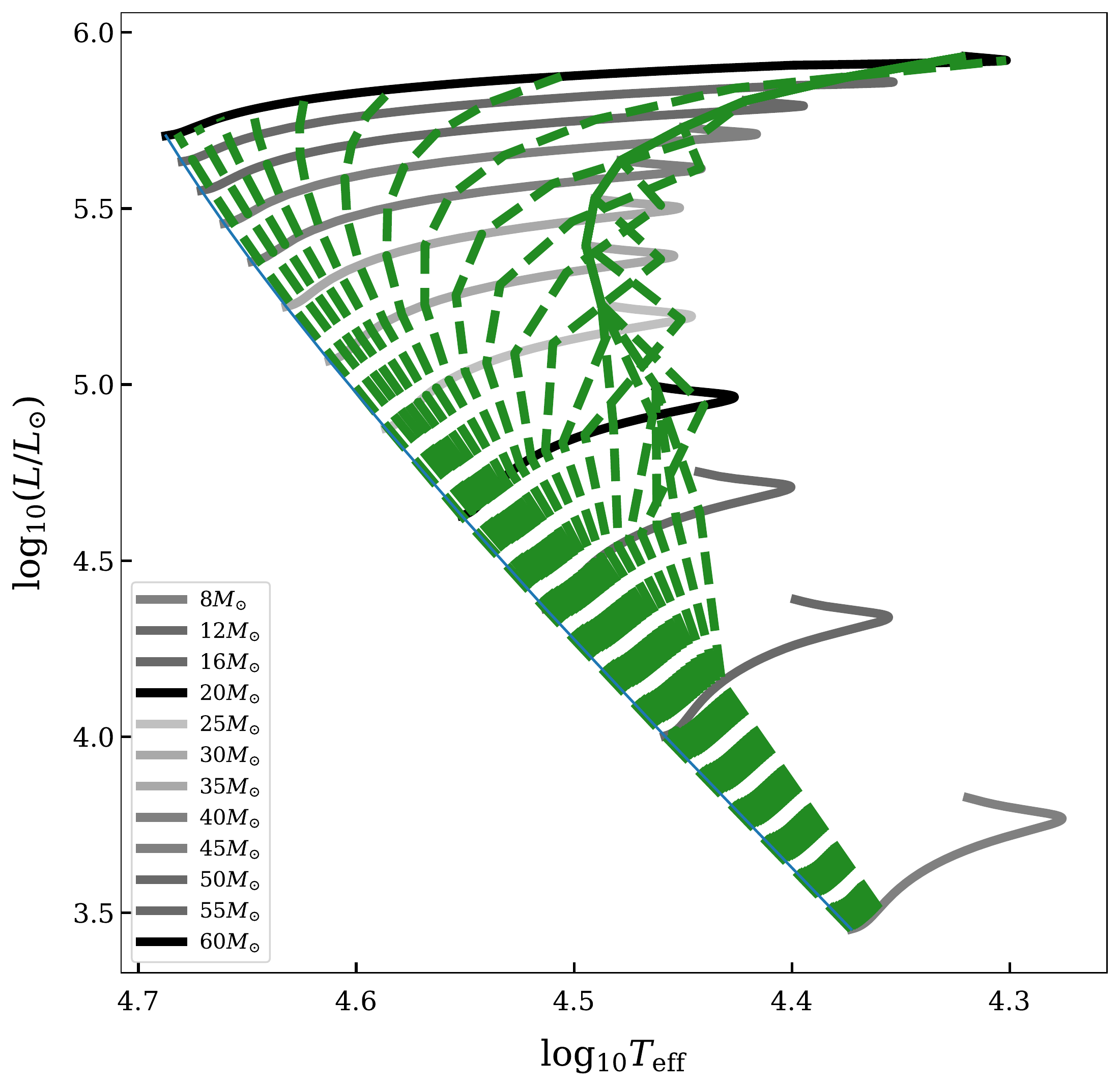}
    \caption{Hertzsprung-Russell diagram of theoretical models calculated for a range of initial masses (8-60\Mdot). Evolutionary tracks are shown in grey, with isochrones displayed in steps of 0.5 Myr from ZAMS until 10 Myr with green dashed lines.}
    \label{fig:HRD_eg_isochrones}
\end{figure}
% \begin{figure}
%     \includegraphics[width = \columnwidth]{ML_ISO_data_1.pdf}
%     \caption{Evolutionary models are presented in the Mass-Luminosity Plane, shown by solid grey lines, with initial masses of 8-60\Mdot\ and with initial rotation rates of 10\% critical and \aov $=$ 0.1. Isochrones are shown in green dashed lines, calculated from ZAMS until 10 Myr in steps of 0.5 Myr.}
%     \label{fig:iso_models}
% \end{figure}
\subsection{Stellar age determination in the M-L plane}\label{agesect}
The isochrone fitting of evolutionary tracks in HRDs has been the dominant method of predicting the age of stars for decades. Due to the observational constraints from spectroscopic analysis and comparisons in colour-magnitude diagrams, this method has been useful for predicting the age of clusters and particularly for low mass stars. However, with large uncertainties in the observed $T_{\rm{eff}}$ and luminosities at higher masses ($>$ 10\Mdot), the related uncertainties in the single isochrone fitting method can be significant. In fact, the inherent systematic uncertainties in using this method -- predicting the age of an observational sample with just 1 grid of evolution models -- has many more consequences. The isochrone method relies on a selected model grid which implements 1 set of input physics, encompassing all internal mixing processes, implying that all stars should evolve with the same amount of convective and rotational mixing regardless of initial mass, metallicity or evolutionary stage. Furthermore, the grid of models chosen for fitting to observations can be selected from a wide range of studies with various codes each implementing physical processes differently, and with varied levels of efficiency. For example, the \texttt{BONN} models by \cite{Bonn11} are calibrated with a 16\Mdot\ model, finding a MS-width corresponding to an \aov\ $=$ 0.335 with a range of rotation rates (0-540\kms) applied for their grid in the mass range of 5-60\Mdot. On the other hand, the \texttt{GENEC} models by \cite{Gen12} fix the amount of internal mixing based on a 1.7\Mdot\ model concluding that the MS-width for stars above 1.7\Mdot\ is reproduced by an \aov $=$ 0.1, with rotation rates set to 40\% critical rotation, which will vary with initial mass. These 2 sets of model grids have been widely-used in comparisons with observations to derive the current age and evolutionary status of O and B stars. Yet depending on which grid of models is selected, the MS-width invoked and the corresponding age estimated for each star will be different. 

Recent work by \cite{clarettorres16, scott21, Jermyn22, Anders22} demonstrates that stars with different initial masses and different evolutionary stages should have different levels of interior mixing. Consequently, the method of determining stellar age has become increasingly important as we develop our knowledge of stellar structure and internal mixing profiles via asteroseismology \citep{moravveji15, bowman20, aerts21} and detached eclipsing binaries. In this work, we develop a more reliable method of predicting the age of individual stars by initially fitting each observation to a theoretical model with specific inputs for convective overshooting and rotational mixing which are calibrated to the observed $T_{\rm {eff}}$, luminosity, and mass of the star. 

\begin{table*}
    \centering
    \begin{tabular}{|c|c|c|c|c|c|c|c|c|}
    \hline\\
    VFTS &  $M_{\rm{spec}}$\ (\Mdot)  &  $M_{\rm{evol}}$\ (\Mdot) & $T_{\rm{eff}}$\ (\rm{kK}) & $\mathrm{log{_{10}}}\;(L/\rm{L}_\odot)$ & \vsini\ \ (\kms) & $t_{\rm{iso}}$\ (\rm{Myr}) & $t_{\rm{iso+}}$ \ (\rm{Myr}) & $t_{\rm{ML}}$\ (\rm{Myr})  \\
      \hline\hline\\
    151  & 139.0  & 53.0 & 37.65 & 5.87 & 118 & 2.750 & 2.875 & 2.841\\
    518  & 22.6  & 47.8 & 44.85 & 5.67 & 112 & 1.500 & 1.500 & 1.917\\
   306  & 18.0  & 28.8 & 31.50 & 5.36 & 90 & 5.250 & 5.750 & 5.643\\
   546  &  10.4  & 19.4 & 31.60 & 4.94 & 94 & 6.250 & 7.750 & 7.668\\
    035 & 16.3  & 16.2 & 32.55 & 4.37 & 346 & 0.375 & 0.500 & 0.478\\
    109 & 9.7 & 11.0 & 24.35 & 4.25 & 352 & 15.25 & 17.87 & 14.81\\

       \hline\\
    \end{tabular}
    \caption{Stellar parameters of O supergiants selected from \citet{Ramirez13} for a range of representative initial masses and ages, including spectroscopic and evolutionary mass estimates. We provide estimates of stellar age with 2 methods, via calibrated evolutionary tracks in the M-L plane, and isochronal ages with \aov\ $=$ 0.1 (iso) and 0.5 (iso+). }%\textb{*Why were they selected? Why only 5? Add VFTS numbers}}
    \label{tab:iso}
\end{table*}
Figure \ref{fig:HRD_eg_isochrones} demonstrates a typical isochrone fitting based on a grid of models for initial masses of 8-60\Mdot\ with \aov$=$ 0.1 and rotation rates of 10\% critical rotation, for ages up to 10 Myr from ZAMS in steps of 0.5 Myr. At the highest mass range ($\sim$ 40-60\Mdot) the systematic error in fitting an observation to 1 isochrone may be on the order of 0.5-1 Myr, compared to the lower mass models ($\sim$ 8-20\Mdot) where a small uncertainty in $T_{\rm {eff}}$ ($\sim$ 0.01 dex) corresponds to an error of up to 10 Myr in isochrones. This factor of 10 difference is partly due to the bending of the MS-band at higher masses towards cooler $T_{\rm {eff}}$ meaning the relative change in age with $T_{\rm {eff}}$ is small. However, this reduced error at high masses is mainly an artefact due to the significantly shorter ($<$ 2 Myr) MS lifetimes of more massive stars above $\sim$ 40\Mdot.

Isochrones can also be fitted to evolutionary tracks in the M-L plane for a given grid of models. We can exclude the uncertainties in observed $T_{\rm {eff}}$ by presenting theoretical isochrones in the M-L plane. Moreover, in \cite{Higgins} the M-L plane was presented as a theoretical tool which can disentangle the effects of internal mixing and wind mass loss, suggesting that we can better compare observations with dedicated models which include the appropriate amount of internal mixing via overshooting and rotation. 
% In any case, fitting an individual star to a theoretical model wi

\begin{figure}
    \includegraphics[width = \columnwidth]{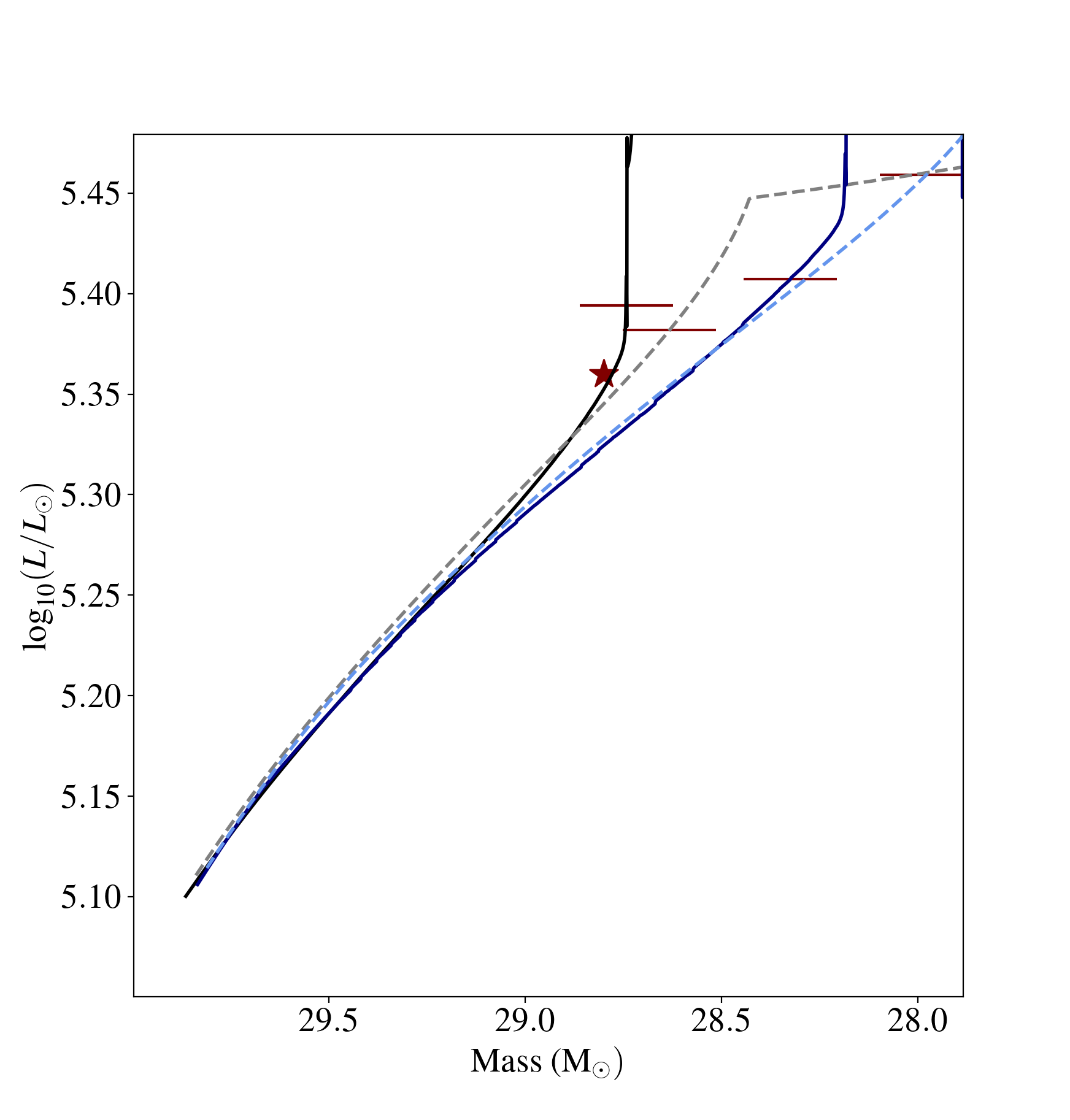}
    \caption{Comparison of VFTS 306 (red star) with $M_{\rm{evol}}$ $=$ 28.8\Mdot\ from \citet{Ramirez13} with evolutionary tracks of 30\Mdot\ models with \aov $=$ 0.1 (solid) and 0.5 (dashed), $\Omega/\Omega_{\rm{crit}}$ $=$ 10\% (black/grey) and 40\% (blue). The horizontal red markers represent the observed log $T_{\rm {eff}}$ $=$ 4.498 of VFTS306 as it is reached for each model.}
    \label{Mevol_eg}
\end{figure}

Our method of reproducing the evolution of an observed massive star and subsequently estimating the current age of the object is as follows. An initial mass is predicted based on the relative position of the ZAMS point on the edge of the forbidden region, shown by the dotted line at the edge of the red shaded region in Fig. \ref{age}, coupled with the gradient of the vector set by the mass-loss rate at the given Z, reaching the observation at a point along the linear vector. 
A theoretical model is then calculated with this initial mass and the standard mass-loss rate from the appropriate wind prescription, which in the case of O-stars would be the \cite{Vink01} rates. 

The model set up also accounts for a default internal mixing prescription which in the first instance includes minimal mixing such as \aov $=$ 0.1 and $\Omega$/$\Omega_{\rm crit}$ $=$ 0.1, which will be modified to fit the observation in the next stage. If the observational data, however, can provide insights on the rotation rate from surface N abundances and \vsini\ then we can already account for this by altering the models rotation rate to match that of the observation (accounting for spin down). 

With our default model set up, we now evolve the model until the observed mass and luminosity are reached. At this point a comparison to the observed $T_{\rm {eff}}$ is needed. If the model's $T_{\rm {eff}}$ is lower than that of the observed $T_{\rm {eff}}$ then the vector length can be extended to reach the higher $T_{\rm {eff}}$ with increased mixing, by rotation or core overshooting. As mentioned, if the rotation rate has already been correlated by the observed \vsini\ or surface abundances, then we can assume that additional mixing by core overshooting is required. We then employ a higher \aov\ of 0.3-0.5 to recalibrate the model in order to reach the observed $T_{\rm {eff}}$ at the same point that the observed mass and luminosity are reached (see Fig.\ \ref{Mevol_eg}).

When the calibrated evolutionary model now reaches the observed $T_{\rm {eff}}$, the model age corresponds to the current age of the observation. This comparison of observed stellar parameters in the M-L plane enables a more robust age estimation from our informed theoretical model.
The M-L plane offers insight into many physical processes acting on the evolution of a star, yet it also provides information on the stage of evolution such as whether a star is on the MS or beyond. While single isochrones estimate the age of a star based on its position in the HRD, in this study we have developed a robust method of estimating the age of stars which excludes standardisation of models. Since isochronal ages are calculated by interpolating between a grid of models which employ a uniform set of input parameters, large discrepancies in MS-length can lead to systematic errors in determining the age of O and B supergiants.

The M-L plane demonstrates that with constant internal mixing (by rotation and convection), then the effective temperatures along the M-L vector also represent isochrones, see Fig. \ref{age}. However, for 5 stars with different rotational and convective overshooting efficiencies, the same age may be reached at different points in the M-L plane. As such, the HRD would look like an age spread, suggesting that isochrones are incorrect as they rely fundamentally on all stars having the same internal mixing parameters. Since we know that stars will have different mixing efficiencies like higher rotation rates or larger core overshooting regions, it seems inaccurate to estimate the age of stars based on a stellar model grid which employs a standardised mixing efficiency across all masses, evolutionary stages and varied Z populations. 

Figure \ref{Mevol_eg} showcases our new method for VFTS 306 with an evolutionary mass of 28.8\Mdot\ alongside a range of model comparisons. This example highlights that depending on the standard set up of each model grid, i.e. with initial rotation rates of 10 \% or 40\% critical, or with enhanced core overshooting compared with minimal overshooting, that the age determined for a particular observation can vary by $\sim$ 1 Myr. We determine the age of each observation with the method as seen in Fig. \ref{Mevol_eg}, where the mass and luminosity are simultaneously reproduced alongside the $T_{\rm {eff}}$. We compare the horizontal markers (log $T_{\rm {eff}}$ $=$ 4.4-4.7) in each evolutionary track with the current observed $T_{\rm {eff}}$ and determine which set of inputs is most representative of the current observables and estimate the age from this model. For VFTS 306 (Fig.\,\ref{Mevol_eg}), the initial mass was 30\Mdot\ with a rotation rate of $\Omega/\Omega_{\rm{crit}}$ $=$ 0.1 and \aov\ $=$ 0.5, with an estimated current age of 5.6 Myr.

% \begin{figure*}
%     \includegraphics[width = 15cm]{ML_Age_Osg_zoom.png}
%     \caption{Evolutionary models are presented in the Mass-Luminosity Plane, shown by solid black lines, with initial masses of 8-60\Mdot\ and with initial rotation rates of 10\% critical. These models are compared with a sample of O supergiants from \cite{Ramirez13}, illustrated by blue circles. The rotation rates of the O supergiant sample are represented by the size of each data point, with increased marker size representing higher \vsini\ values (normalised for 0-600\kms). We illustrate 4 example log T$_{\rm{eff}}$ values of 4.4-4.7 in steps of 0.1 dex with coloured horizontal lines (red, blue, orange and green) for each model, which are used as a comparison to the observed log T$_{\rm{eff}}$ of each O supergiant. We include a closer look at the 16\Mdot\ and 20\Mdot\ tracks with observations of varying \vsini\ to illustrate our method.}
%     \label{fig:ML_Age_Osg_vsini}
% \end{figure*}

Our method of calibrating 5 stellar parameters (3 observed quantities and 2 physical inputs which are calibrated) in evolutionary models via the M-L plane allows for accurate estimates the age of stars when the observed luminosity, mass and effective temperature are reached simultaneously, in line with our estimates of rotation and overshooting. This may provide a more reliable method of estimating the age of stars, even for a large sample.% as illustrated in figure \ref{fig:ML_Age_Osg_vsini}. 

Figure \ref{age} illustrates an extension of the M-L plane from \cite{Higgins} by estimating the age of an object where it's observed effective temperature is reached in the M-L plane and corresponds to a calibrated evolutionary model, for a given rotation rate and \aov. This is showcased by example temperature markers added along the evolutionary model's vector shown in Fig. \ref{age} as log $T_{\rm {eff}}$ $=$ 4.4 and 4.3, to highlight that as the model evolves with time towards cooler effective temperatures, we can compare the model $T_{\rm {eff}}$ to that of the observation.

 %VFTS 500 and VFTS 642 provide an opportunity as a calibrated test-bed for this method of determining stellar ages since the system has been reproduced with evolutionary models. We note the age at which these models concurrently reach the observed stellar parameters is 2.9 $\pm$ 0.1 Myrs, which lies in agreement with \citet{Taylor}.

% \begin{figure}
%     \includegraphics[width = \columnwidth]{test_iso.png}
%     \caption{Stellar models are shown in the Mass-Luminosity Plane, illustrated by solid grey lines, with initial masses of 8-60\Mdot\ and with initial rotation rates of 10\% critical. These models are compared with a sample of O supergiants from \cite{Ramirez13}, shown by coloured triangles. The rotation rates of the O supergiant sample are represented by the colour of each data point, with \vsini\ values shown in the colorbar. We provide isochrones of our model grid in dashed coloured lines ranging from 0.5-2.5Myrs.}
%     \label{fig:iso_models}
% \end{figure}

\begin{figure}
    \includegraphics[width = \columnwidth]{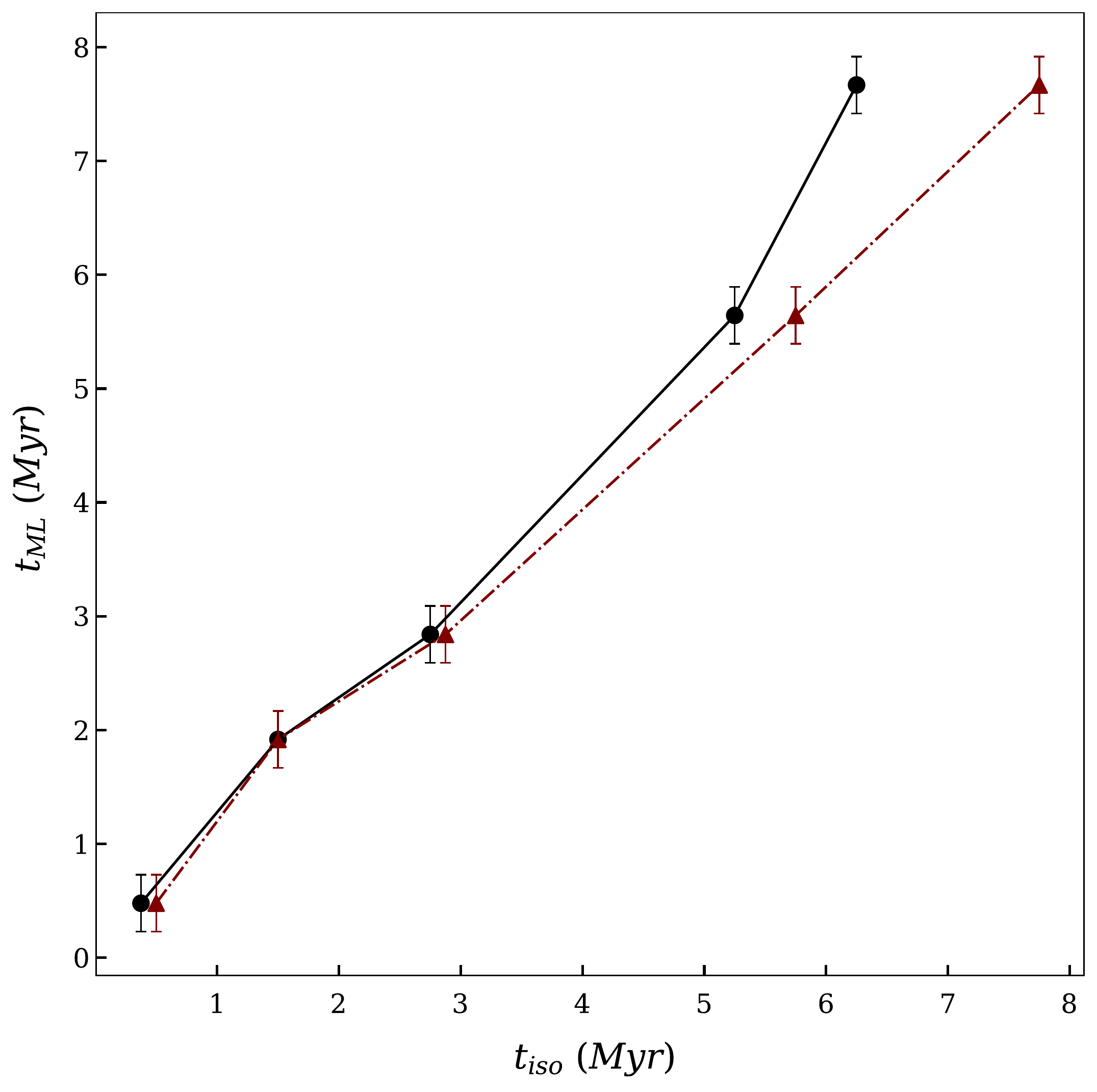}
    \caption{Stellar age of 5 O supergiants selected from the VFTS sample to represent a range of initial masses and current ages, calculated in the M-L plane and compared with the ages inferred from isochrones with 2 assumptions for \aov $=$ 0.1 (black solid) and 0.5 (red dashed).}
    \label{Agecomparison}
\end{figure}
\begin{figure}
    \includegraphics[width = 8cm]{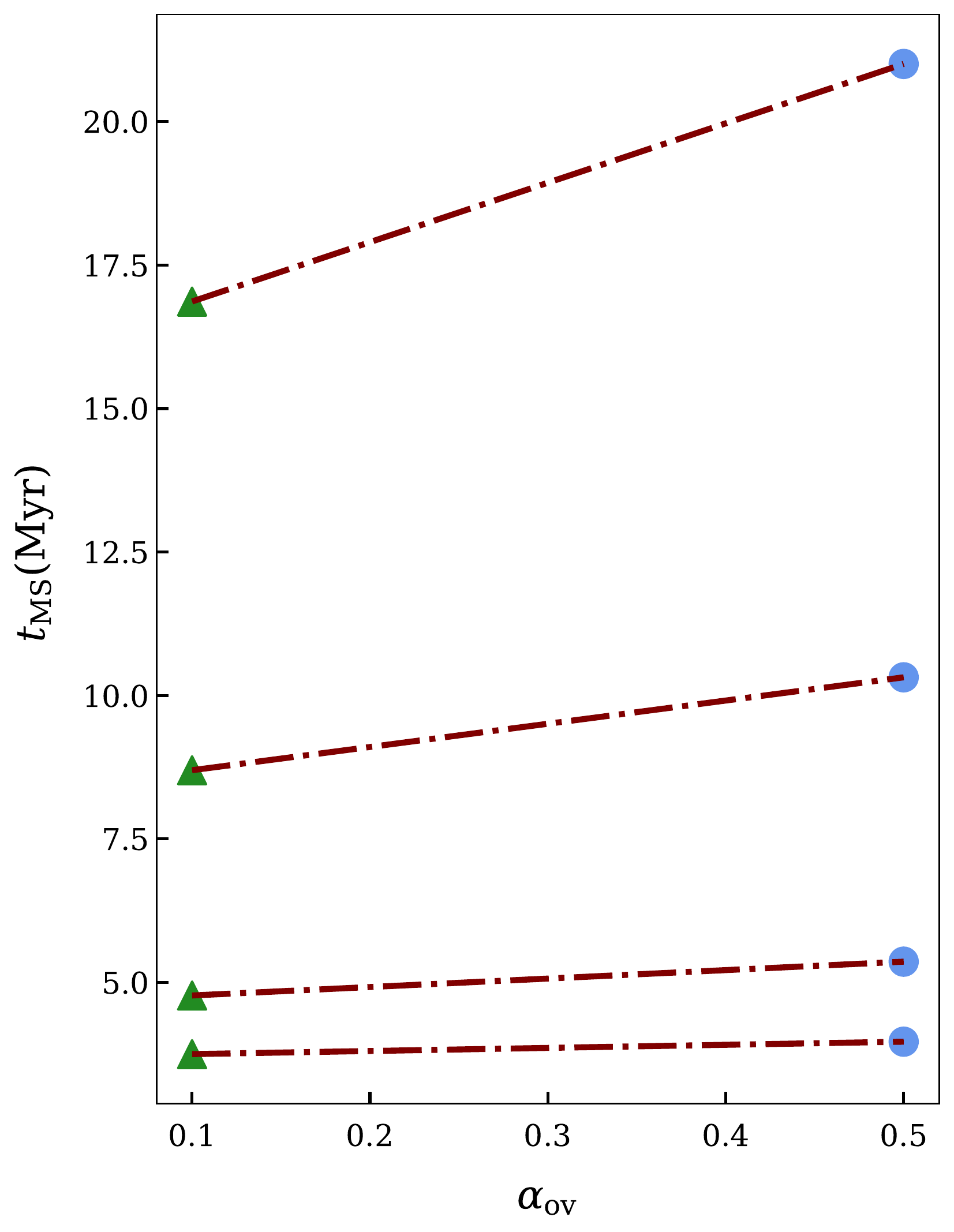}
    \caption{Stellar age as a function of \aov\ of 4 models with representative initial masses of 12, 20, 40, and 60\Mdot\ with 2 assumptions for \aov $=$ 0.1 (green triangles) and 0.5 (blue circles). The red dashed line shows the difference in MS-lifetime when implementing different values of \aov. Lower initial mass models have the longest MS length}
    \label{t_MS_ov}
\end{figure}
\subsubsection{O supergiant age estimates}\label{VFTSage}
The first step in determining an age constraint of observed stars with the M-L plane method is to estimate the initial mass based on gradient and M-L relation on the ZAMS. Then we compare the relevant observed \vsini\ with our rotating model grid such that fast rotators (\vsini\ > 200\kms) are compared with 40\% critically rotating models and slow rotators (\vsini\ < 200\kms) are compared with 10\% critically rotating models. Of course, in theory the model grid can be refined to include more detailed sets of parameters which explore the entire range of possible model parameters. The effective temperature of the calibrated model and observed $T_{\rm {eff}}$ are then compared. The stellar model is finally corrected for the appropriate amount of internal mixing in order to better reproduce the $T_{\rm {eff}}$ in M-L space. This means that the observed mass, luminosity and $T_{\rm {eff}}$ are reached simultaneously, where the now mixing-corrected model reaches the $T_{\rm {eff}}$ of the observed star, we can establish an age constraint with a given uncertainty.
% Figure \ref{fig:ML_Age_Osg_vsini} represents rotating models of $\Omega$/$\Omega_{\rm crit}$ $=$ 0.1 for comparison with O supergiants categorised by their associated marker size (normalised for 0-600\kms). For instance, a `blue' coloured data point from Figs. \ref{01datacolor}-\ref{04datacolor} would be slowly rotating and should be compared with evolutionary tracks from Fig. \ref{01datacolor} since these are slowly rotating models with $\Omega$/$\Omega_{\rm crit}$ $=$ 0.1. On the other hand, a `red/yellow' coloured data point from Figs. \ref{01datacolor}-\ref{04datacolor} would be a fast rotator and should be compared with evolutionary tracks showcased in Fig. \ref{04datacolor}, since they are calculated with $\Omega$/$\Omega_{\rm crit}$ $=$ 0.4.

The M-L plane also provides a new method of determining the age of a large sample of observations by comparing with a calibrated grid of evolutionary models.  The effective temperatures of our models can then be used to equal the observed effective temperature of each data point. This point is then used to provide an estimate of the age at that temperature range in the evolutionary model. For a given observation the current mass, luminosity and effective temperature can be reproduced by calibrating the comparative evolutionary model based on the \vsini\ and extra mixing required via \aov. 

In each case, the chosen grid of models can be measured alongside the given data point until the observed mass and luminosity are reached. At this point the observed $T_{\rm {eff}}$ is compared with the evolutionary track's $T_{\rm {eff}}$ to check if the data point's mass and luminosity is reached at the same point as the observed $T_{\rm {eff}}$. If the model's current $T_{\rm {eff}}$ is lower than the observed $T_{\rm {eff}}$ then the observation should be compared with the grid of evolutionary tracks which accounts for additional mixing by \aov $=$ 0.5. Now, when the track reaches the observed $T_{\rm {eff}}$, we can estimate the age of this given data point, based on its calibrated rotation rate and core overshooting \aov.

Table \ref{tab:iso} demonstrates the range in age estimates from our model grid with 2 methods. For a set of 5 representative O supergiants (selected for a range of initial masses and current ages) we calculate the age with the standard isochrone method, but for 2 values of \aov\ $=$ 0.1 and 0.5. This highlights that for stars in the mass range 20-30\Mdot\ which have MS lifetimes of $\sim$ 5-10 Myr that systematic uncertainties from default values of \aov\ may correspond to an error of 10\% of the $t_{\rm{MS}}$. On the other hand, estimating the age in the M-L plane allows for direct calibration of rotation rates and \aov\ based on stellar observables. In this case, the uncertainties depend on analysis of individual stars and their model comparisons, such as errors in $M_{*}$, $L_{*}$ or grid refinement. 
\begin{figure}
    \includegraphics[width = \columnwidth]{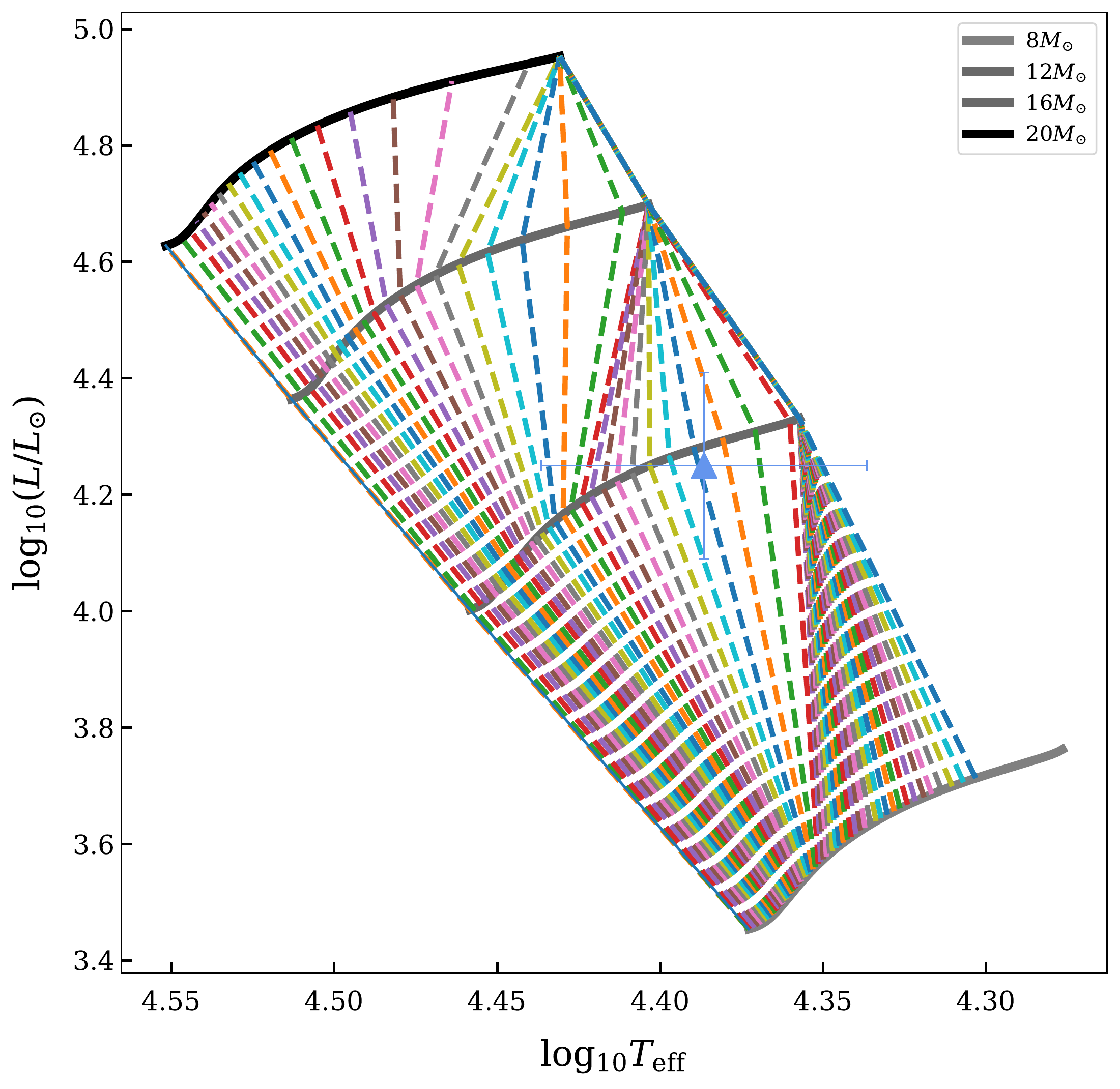}
    \caption{Comparison of VFTS 109 (blue triangle) with $M_{\rm{evol}}$ $=$ 11.0\Mdot\ from \citet{Ramirez13} with evolutionary tracks of 8-20\Mdot\ with \aov $=$ 0.1 and $\Omega/\Omega_{\rm{crit}}$ $=$ 10\%. The coloured lines represent isochrones for 0-30 Myr in steps of 0.5 Myr.}
    \label{VFTS109}
\end{figure}
\begin{figure}
    \includegraphics[width = \columnwidth]{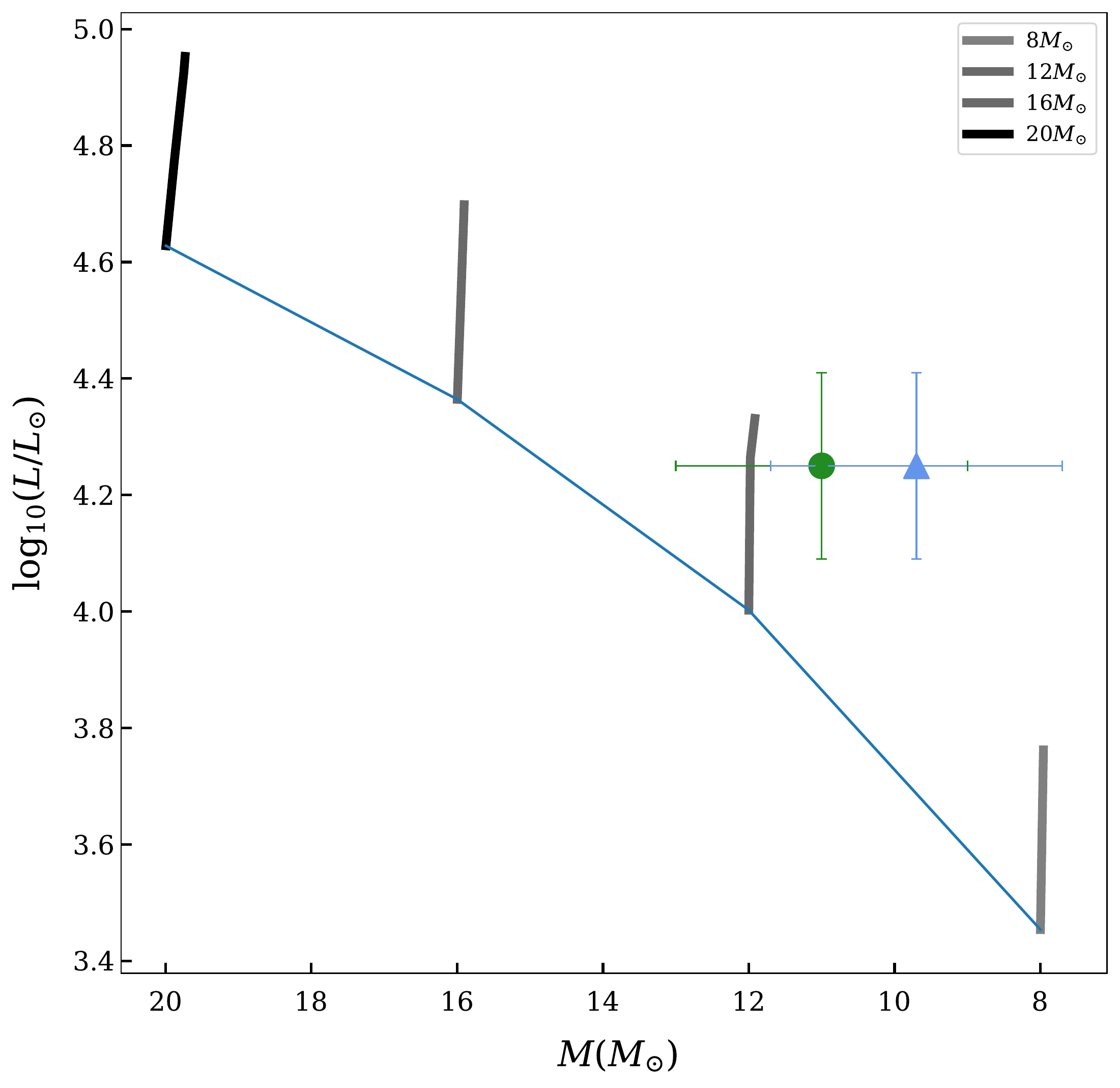}
    \caption{Mass-Luminosity Plane showing VFTS 109 with $M_{\rm{evol}}$ $=$ 11.0\Mdot\ (green circle) and $M_{\rm{spec}}$ $=$ 9.7\Mdot\ (blue triangle). Evolutionary tracks of 8-20\Mdot\ are shown in grey with \aov $=$ 0.1 and $\Omega/\Omega_{\rm{crit}}$ $=$ 10\%. The blue coloured line represent the ZAMS.}
    \label{VFTS109_ML}
\end{figure}

We test the effectiveness of our new method in the M-L plane alongside the isochronal method in Fig.\,\ref{Agecomparison}. We compare the estimates of age for each of the 5 O supergiants shown in Table \ref{tab:iso}. The estimates from both methods are in good agreement, particularly when including the model grid adopting core overshooting \aov\ $=$ 0.5 ($t_{\rm{iso+}}$). However, isochronal ages underestimate the age of lower mass objects, for example by 1.4 Myr for VFTS 546 ($\sim$ 20\Mdot). In fact, for lower mass observations $\leq$ 20\Mdot\ the systematic uncertainties can be significant. We show in figure \ref{t_MS_ov} that lower mass stars with longer MS lifetimes will have larger uncertainties due to \aov\ than higher mass stars. While the relative uncertainty for $M_{\rm{init}}$ > 30\Mdot\ is only 5-20\%, these uncertainties can be as high as 20-25\% for $M_{\rm{init}}$ < 30\Mdot.

In this paper we first demonstrate the method in the M-L plane with Fig.\ref{age}. We then provide an example of calibrating each observation to a refined evolutionary model which reaches the observed $T_{\rm {eff}}$ at the correct M-L location, i.e. for the observed M and L. We provide a detailed comparison of our 4 model grids to 1 O supergiant in the intermediate mass range $\sim$30\Mdot\ in Fig.\ref{Mevol_eg}. Our model grids are designed to highlight extremely high and low rotation, and high and low overshooting. Of course, in theory the model grid can be refined to include more detailed sets of parameters which explore the entire range of possible model parameters. Since the O star range is likely represented largely by 0$-$40\% critically rotating models \citep{Ramirez13} and \aov$=$0.1$-$0.5, we adopt this set up for our comparisons. Fig. \ref{Mevol_eg} demonstrates that a given $T_{\rm {eff}}$ varies widely for each of the 4 models depending on which extreme set of mixing parameters is selected. Moreover, the point at which a $T_{\rm {eff}}$ is reached in M-L space, varies depending on the selection of $\Omega/\Omega_{\rm{crit}}$ and \aov, with constant initial mass and Z. We show that the model which lies closest to the observed O supergiant (VFTS 306) best represents the mixing and mass loss of the observation. The uncertainty in the star's calibrated model age will depend on the inherent $T_{\rm {eff}}$ uncertainty from \cite{Ramirez13} (for example log $T_{\rm {eff}}$ $=$ 4.498 $\pm$  0.015, we compare the $\Delta$ $t$ at log $T_{\rm {eff}}$ $=$ 4.514 giving an uncertainty of  $\sim$ 0.1 Myr).

We compare the stellar ages calculated in the M-L plane alongside isochronal ages for 5 O stars. While we find our method lies in good agreement with the standard isochrone method, we find that lower mass stars $\sim$ 20\Mdot\ can have a large uncertainty due to systematic errors in the MS-width adopted for the entire mass grid. The most massive objects are not affected by internal mixing values to the same degree, due to the reduced pressure scale height at higher masses. We show the variation in isochronal age depending on the selection of \aov\ $=$ 0.1 and 0.5. For the case of VFTS 546 ($M_{*}$ $\sim$ 20\Mdot) we find an uncertainty of $t$ $\sim$ 2 Myr for the isochronal age estimates when implementing \aov\ $=$ 0.1 or 0.5.

We provide an analysis of an O star (VFTS 109) with estimates of the current age in a standard HRD using isochrones, in the M-L plane with the same grid of models, and finally with a calibrated model in the M-L plane. We find that when estimating the age of VFTS 109 with standard ischrones from Fig. \ref{VFTS109} in a HRD, there are separate uncertainties related to the observed luminosity and $T_{\rm {eff}}$. We calculate an age of 15.25 Myr with an error of 0.5 Myr due to luminosity, and an error of 15 Myr due to $T_{\rm {eff}}$. 

%In the M-L plane, we compare the same standardised isochrones finding an age of 14.25Myr with an uncertainty of 26Myr.
Finally, we calibrate the evolution of VFTS 109 in the M-L plane by estimating the initial rotation rate to be 40\% critical rotation (with an observed \vsini\ of 352\kms), and core overshooting \aov\ $=$ 0.1. This corrected-model now reaches the observed $T_{\rm {eff}}$ simultaneously in the M-L space with the observed $M$ and $L$ of VFTS 109, at an age of 14.81 Myr. Comparatively, the error due to the initial rotation rate is low (0.1 Myr), while the uncertainty due to \aov\ is 3.1 Myr, with the corresponding model implementing \aov\ $=$ 0.5, providing an age of 17.87 Myr. This exercise highlights the key systematic and empirical uncertainties, where the largest systematic uncertainties lie in the choice of \aov\ since this directly extends the MS-lifetime, while the largest empirical uncertainties are due to the estimated $T_{\rm {eff}}$. In general, the crucial error is assuming the same \aov\ for all stars inducing a significant uncertainty of $\lesssim$ 3 Myr, in addition to the empirical errors. While spectral modelling continues to improve, reducing uncertainties in stellar observables such as $T_{\rm {eff}}$, we can already reduce the theoretical errors due to model inputs by directly calibrating observations with appropriate mixing efficiencies in the M-L plane. Figure \ref{VFTS109_ML} shows both $M_{\rm{evol}}$ and $M_{\rm{spec}}$ for VFTS 109 in the M-L plane to highlight the discrepancy in determining stellar mass from spectral modelling and evolution models. The $M_{\rm{spec}}$ presents VFTS 109 as a post-MS object with large uncertainties ($\sim$ 2\Mdot, see also Sect. \ref{lmcvfts}).  

In this work, we provide an alternative method of ageing individual stars, with an approach which eliminates the largest systematic uncertainties by first allowing calibration of theoretical processes such as core overshooting and rotational mixing, before fitting models to observed stellar parameters. Yet our new method does have a drawback in using stellar mass as a constraint on the evolution, since it is not directly measurable from observations but is inferred from log $g$ in radiative transfer models. This means that stellar mass estimates carry uncertainties based on the method from which it was calculated. Moreover, spectroscopic masses inferred from spectral analysis can vary by up to a factor of 2 from evolutionary masses calculated from evolution models \citep{Herrero}. So while these inconsistencies persist, the mass estimates for stars will remain imprecise. The most robust method will rely on accurate mass estimates from eclipsing binaries where Kepler's laws provide dynamical masses which are highly accurate and reliable \citep[e.g.][]{Burkholder, WV10, Mahy2017}, giving precise age estimates (see Sect. \ref{binaries}). 

% *Discussion of ML plane drawbacks with masses provided spectroscopically and from comparisons with evolution models. Show both in ML plane for isochrones and show that spectroscopic masses are highly uncertain for the O supergiants. Can we use the ML plane to teach us about these drawbacks or which mass estimate is likely more correct? We can compare the full sample of O and B supergiants and see what we can learn from the ML plane. Ultimately the key will be to use binaries with dynamical masses to calibrate the mixing for a wider sample.

\begin{figure}
\centering
\includegraphics[width = \columnwidth]{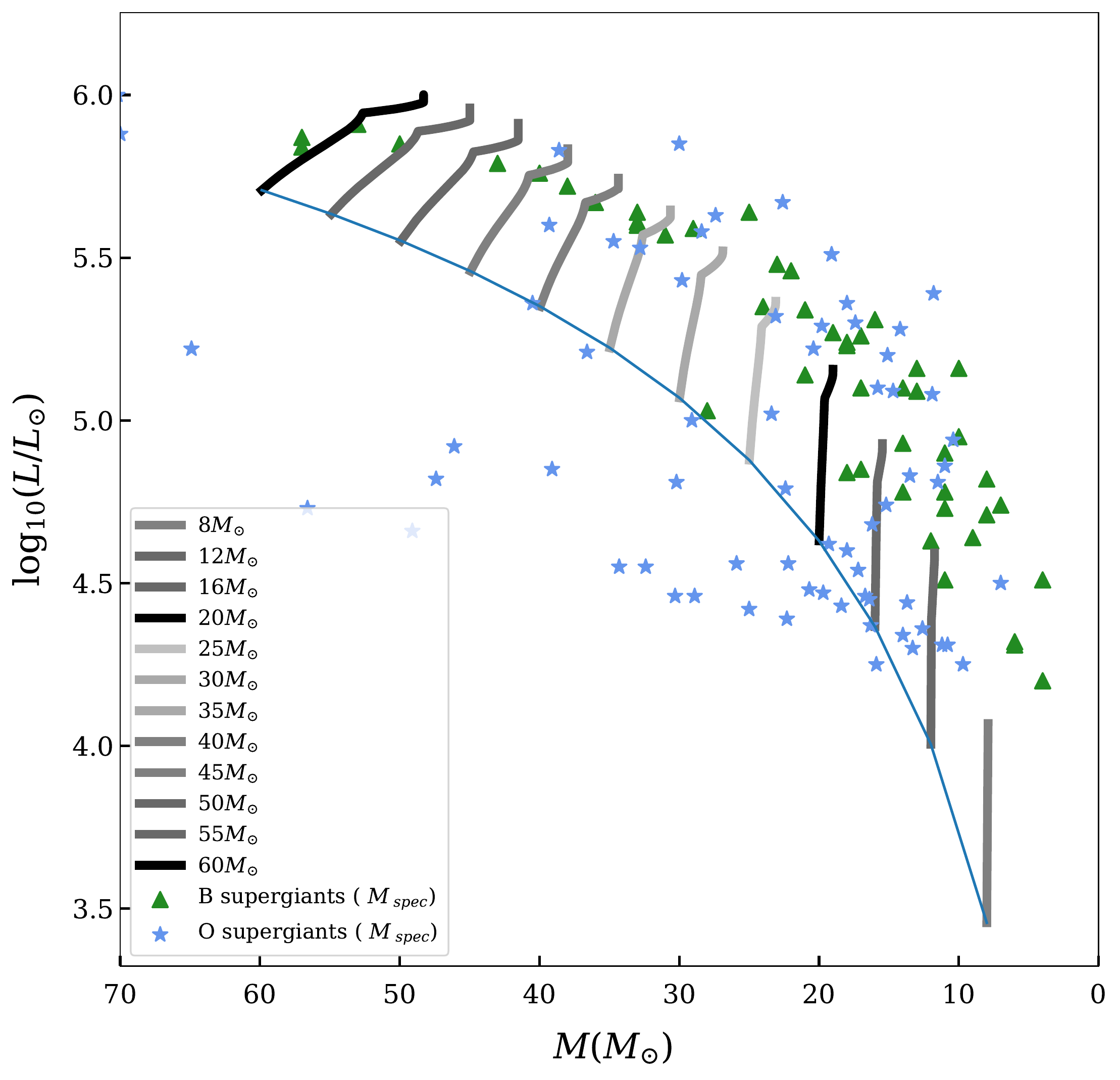}
\caption[Grid of models with O supergiants, O dwarfs and B supergiants ($M_{\rm spec}$)]{\footnotesize Grid of evolutionary models compared in the M-L plane with initial masses 8-60\Mdot\ for \aov\ $=$ 0.1 are shown in grey solid lines, with the ZAMS of each model shown by the blue solid line. Comparisons with O supergiants (blue stars) and B supergiants (green triangles) are shown from \citet{Ramirez13} and \citet{McEvoy} adopting spectroscopic masses.  }
\label{Mspec}
\end{figure}
\begin{figure}
\centering
\includegraphics[width = \columnwidth]{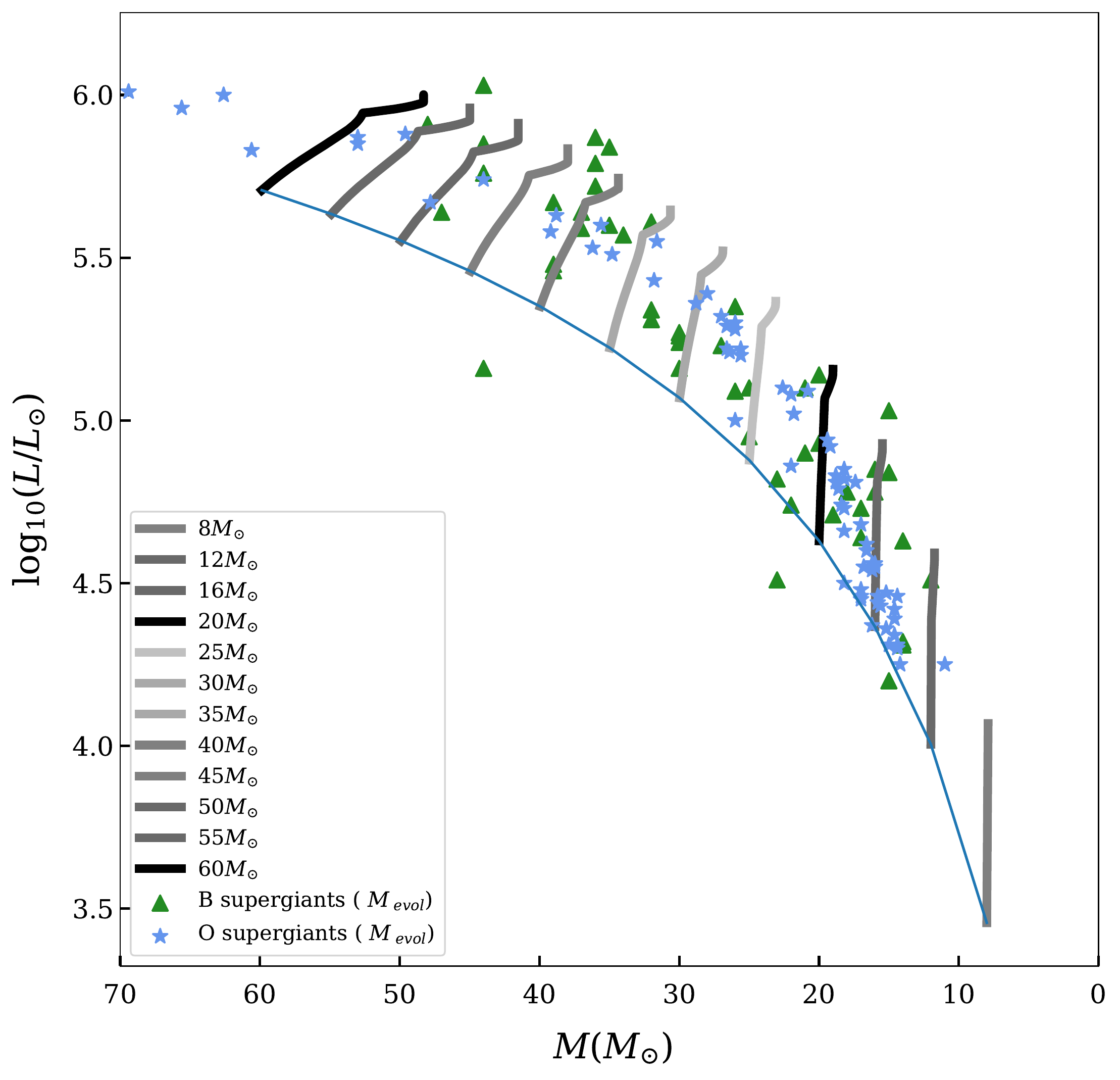}
\caption[Grid of models with O supergiants, O dwarfs and B supergiants ($M_{\rm spec}$)]{\footnotesize Grid of evolutionary models compared in the M-L plane with initial masses 8-60\Mdot\ for \aov\ $=$ 0.1 are shown in grey solid lines, with the ZAMS of each model shown by the blue solid line. Comparisons with O supergiants (blue stars) and B supergiants (green triangles) are shown from \citet{Ramirez13} and \citet{McEvoy} adopting evolutionary masses.  }
\label{Mevol}
\end{figure}
\subsection{Mass discrepancy of O and B supergiants from the VFTS sample}\label{lmcvfts}
For the last few decades, atmosphere modelling of stellar spectra has improved allowing more reliable spectroscopic masses, however with some systematic uncertainties remaining. Code comparisons highlight the remaining challenges in estimating spectroscopic masses at the upper mass range. With these inherent uncertainties in spectral modelling, another method of calculating stellar masses from observations proved a useful alternative. Evolutionary model grids for varying masses allowed comparisons with observations to predict the evolutionary mass of a star at a given age. While this method incorporated many similar systematic uncertainties inherent to theoretical models, it provided a comparison or alternative to spectroscopic masses. Now, the misalignment of these 2 methods in estimating stellar masses of observational samples has led to a widely known problem called the ``Mass discrepancy problem'', whereby predictions from evolution models compared to spectroscopic models can vary from a few solar masses to a factor of 2 difference, leading to a wide range of mass predictions. The most concerning challenge of this discrepancy is the divergence at the upper mass range, which becomes so uncertain that in cases utilising the evolutionary masses becomes futile. Uncertain mass-loss rates and implementation of wind recipes further cause issues in estimating the evolutionary mass of the most massive stars.

In a similar way, the stellar age of massive star observations is calculated by interpolating between various stellar evolutionary tracks, and estimating the age based on unconstrained input physics which are incorporated into stellar models, varying the MS-length, luminosity and many other factors. In this paper we aim to highlight these uncertainties in estimating the spectroscopic and evolutionary masses of massive stars, providing evidence through large samples. Furthermore, we provide a useful tool in calibrating the evolutionary path of a massive star for comparison to observations in order to give a more robust estimate of stellar age, as previously outlined in Sect. \ref{agesect}. While the M-L plane method carries uncertainties in stellar mass from spectroscopic or evolutionary mass estimates, we may use the M-L plane to decipher which method provides the most likely accurate estimates. 

We test the M-L plane method in this section for a wider sample of single O and B supergiants. Due to the nature of the initial mass function (IMF), there are few homogeneous observational samples of massive stars, many of which provide relatively uncertain spectroscopic masses which are increasingly uncertain with increased mass. Owing to their high luminosities, massive stars tend to have upper limits on their spectroscopic masses. Moreover, as stars evolve during their core-H burning stage, they increase in luminosity. This characteristic is applicable in stars of all masses up to approximately 60\Mdot, above which stars may evolve as WNh (H-burning WR-like stars) which are close to their Eddington limit, making it even more challenging to estimate their mass based on their luminosities \citep{Crow10, Sabh22}.

\cite{Ramirez13} provides spectroscopic analysis of the O supergiant sample of the VFTS, with evolutionary masses calculated from comparisons with the \texttt{BONN} evolutionary tracks \citep{Bonn11} and interpolated with the \texttt{BONNSAI} tool \citep{Bonnsai}. \cite{Grin17} finds that the uncertainty in spectroscopic masses of the O supergiant sample is significantly larger than the evolutionary masses and as such utilises the estimates from evolutionary masses in their analysis. Figure \ref{Mspec} shows the error in spectroscopic masses of the O supergiant sample since most of the lower mass objects lie below the `forbidden' region set by the mass-luminosity relation. On the other hand, some O supergiants lie beyond the MS suggesting their mass estimates are systematically too high. Moreover, Fig. \ref{VFTS109_ML} further showcases the uncertainty in O supergiant spectroscopic masses, with VFTS 109 positioned beyond the MS.

We then compare the evolutionary mass estimates for the O and B supergiants in Fig. \ref{Mevol} finding that while implementing the evolutionary masses, the populations now occupy the same evolutionary state, i.e. not in earlier and later evolutionary positions as would be expected in a HRD. However, the B supergiant $M_{\rm{evol}}$ assume an evolutionary stage based on a standard model grid, i.e. if the MS width (based on the selected \aov) does not enclose the B supergiants then they are considered post-MS objects which will have consequences for the inferred evolutionary masses \citep[see Fig.\ref{Mevol} and][]{McEvoy}. 
\begin{figure}
\centering
\includegraphics[width = \columnwidth]{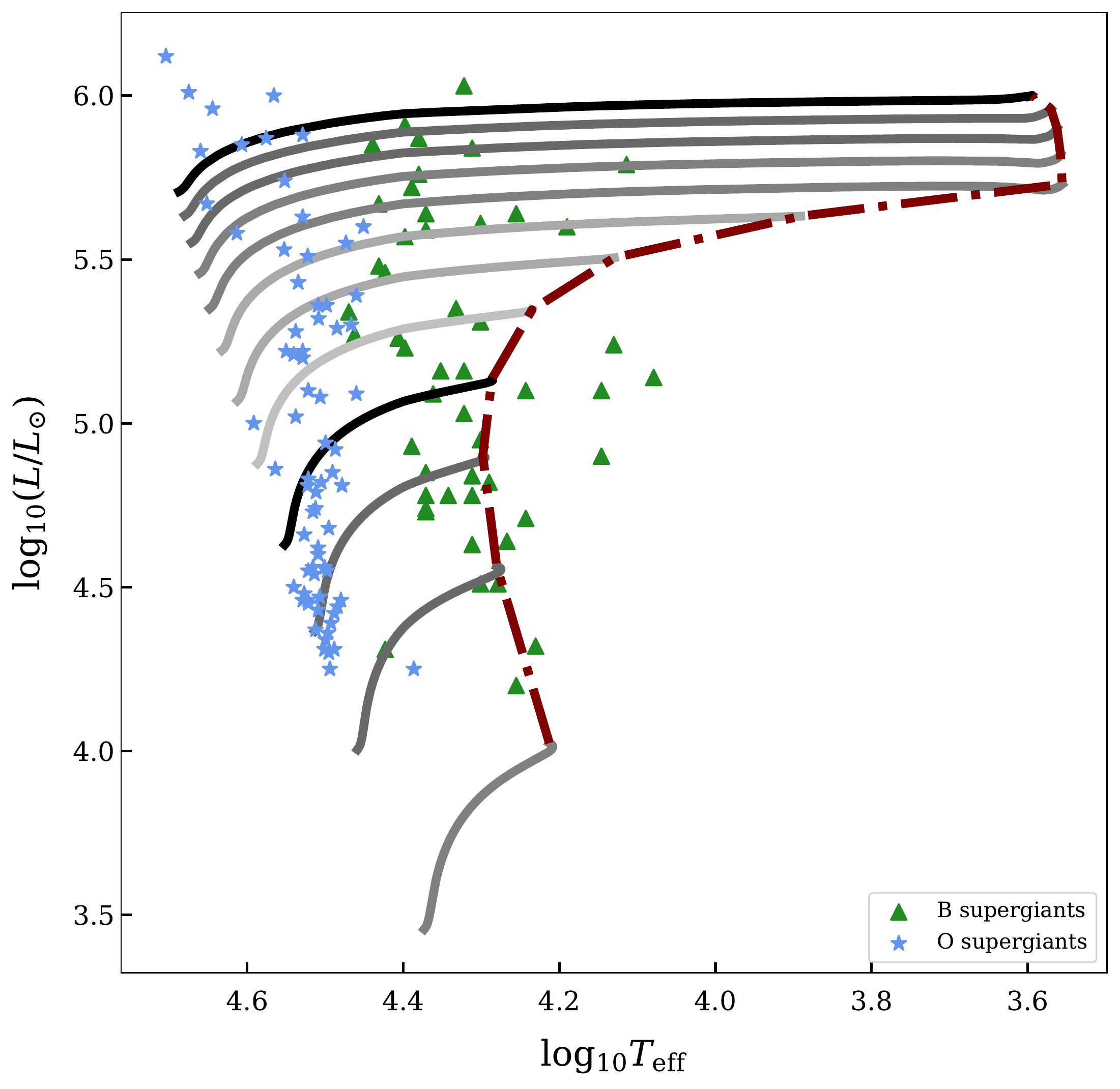}
\caption[Grid of models in HRD with observations for \aov\ $=$ 0.1, 0.5]{\footnotesize Evolutionary tracks of rotating models (10\%\ critical rotation) with initial masses of 8-60\Mdot, for \aov $=$ 0.5 (grey solid lines) where the TAMS location is highlighted by the dashed red line. Observations of O supergiants (blue stars), and B supergiants (green triangles) from \citet{Ramirez13}, and \citet{McEvoy} are shown for comparison.}
\label{HRDOB}
\end{figure}
\begin{figure}
\centering
\includegraphics[width = \columnwidth]{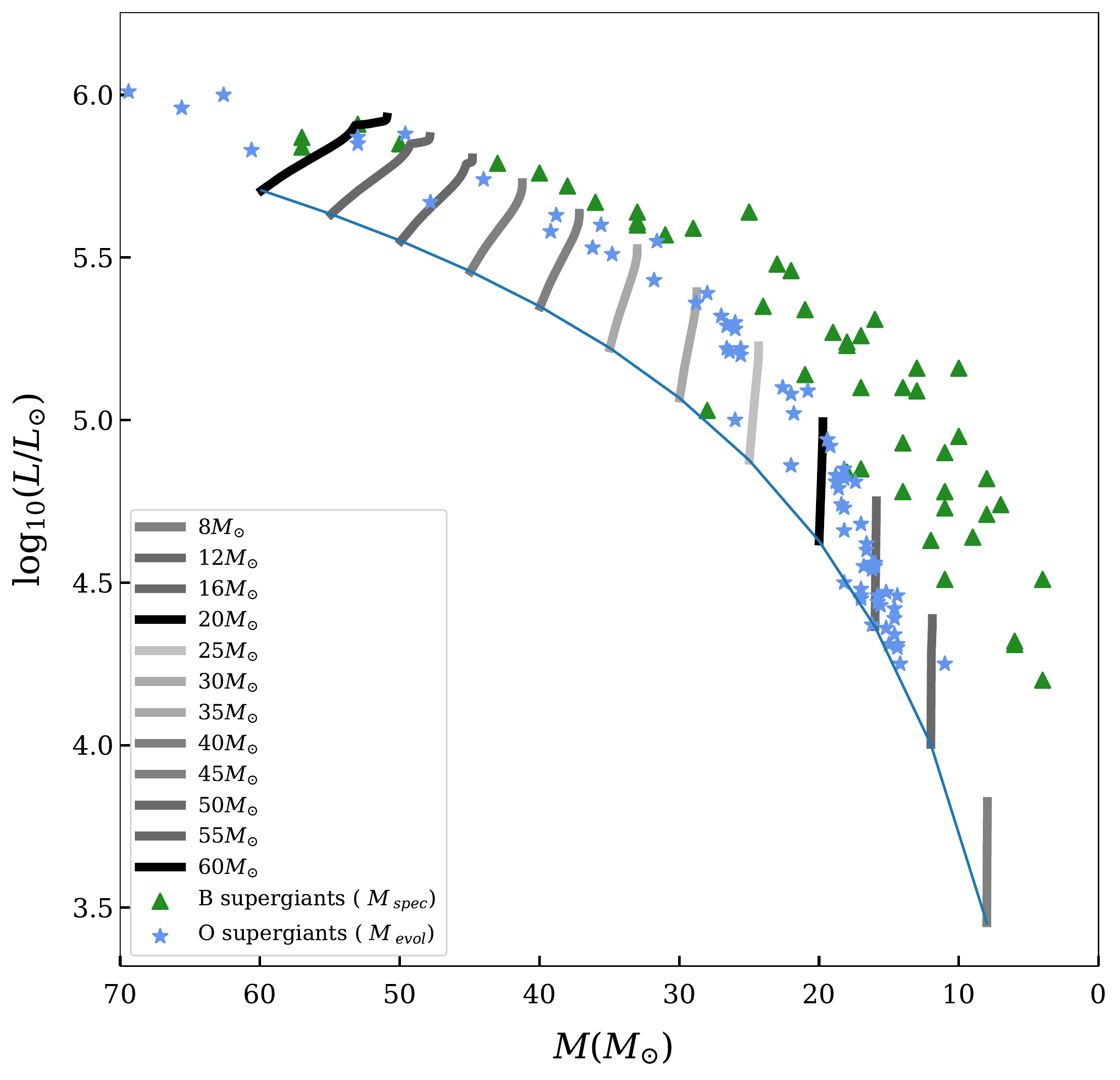}
\caption[Grid of models with O supergiants and B supergiants ($M_{\rm spec}$)]{\footnotesize Grid of evolutionary models compared in the M-L plane with initial masses of 8-60\Mdot\ for \aov\ $=$ 0.1 are shown in grey solid lines, with the ZAMS location highlighted by the solid blue line. Comparisons with O supergiants adopting evolutionary masses (blue stars) and B supergiants adopting spectroscopic masses (green triangles) are shown from \citet{Ramirez13} and \citet{McEvoy}. }
\label{MspecBMevolO}
\end{figure}
In \cite{McEvoy}, the same \cite{Bonn11} evolution tracks were used in estimating the evolutionary masses of the B supergiants, however, in some cases the B supergiants were considered pre-TAMS objects and in others, post-TAMS objects. This assumption is based on the input physics of the \cite{Bonn11} models, where \aov\ is assumed to be 0.335, estimated from calibration of a 16\Mdot\ model alongside the drop in \vsini\ with log g $< 3.2$. If a wider MS is adopted, as suggested by \cite{Vink10} and \cite{McEvoy} then many more of the B supergiant sample are included in the pre-TAMS class than the post-TAMS sample. In fact, from Fig. \ref{HRDOB} with a small increase from \aov\ $=$ 0.335 to 0.5, the number of post-TAMS objects drops to $\sim$6 from a sample of 34 B supergiants. This means that over 80\% of B supergiants could be H-burning objects, while other uncertainties such as errors in log $T_{\rm {eff}}$ are still unaccounted for. An underestimate of $T_{\rm {eff}}$ for the B supergiant sample would further steer the B supergiant sample towards being H-burning objects. 

Considering the timescales of core H- and He-burning, it may be more likely that stars that lie close to the TAMS position are still burning H since the last 1\% of H-burning is approximately 100,000yrs which is comparable to the entire core He-burning timescale. \cite{McEvoy} suggests a representative random error of 1000K in $T_{\rm {eff}}$ estimated for all B supergiants due to fitting procedures but also highlights that there may be additional systematic uncertainties not accounted for. An uncertainty or underestimate in log $T_{\rm {eff}}$ $\sim$ 0.2dex would already account for the remaining 6 outliers of the post-TAMS B supergiants.

In \cite{Higgins}, we find that our test-bed required extra internal mixing in order to reproduce observed luminosities by enhanced core overshooting \aov\ $=$ 0.5. If this conclusion is applied to our LMC grid of models, we must consider B supergiants as core H-burning MS objects. \cite{Vink10} considers the effects of bi-stability braking as a method of reproducing the sample of slow-rotating B supergiants from the VLT-FLAMES sample \cite{Evans05}, enabling them to be categorised as MS objects which have not spun-down over their MS lifetime. Figure \ref{HRDOB} demonstrates that B supergiants may be included in MS evolution with \aov\ $=$ 0.5. Due to an unbiased data set from VFTS, we do not observe a gap between O and B supergiants, suggesting B supergiants may in fact prove to be H-burning objects which lie close to the TAMS.

The M-L plane has illustrated a discrepancy between the evolutionary and spectroscopic masses of B supergiants from the VFTS sample. When comparing the O supergiants and B supergiants in the M-L plane implementing the evolutionary masses of the B supergiant sample, they tend to populate the same spectral class as the O supergiant region suggesting that they are not more evolved than the O supergiant sample, as would be expected in a HRD. This suggests an overestimate in the evolutionary mass predictions of the B supergiant sample, or that considering B supergiants as post-MS objects is incorrect. We find that when implementing a widened MS, as previously suggested by \cite{Vink10} and \cite{McEvoy}, by including \aov\ $=$ 0.5, that over 80\%\ of the B supergiant sample now lie within the MS and should be considered H-burning objects.

Since B supergiants are expected to lie beyond O supergiants, we compared B supergiants with spectroscopic masses from \cite{McEvoy} rather than evolutionary masses finding that indeed the B supergiants now occupied a later evolutionary phase in the M-L plane (see Fig.\, \ref{MspecBMevolO}). This suggests that the discrepancy between evolutionary and spectroscopic masses is consequential for B supergiants and their evolutionary stage. In this study we find a systematic overestimation of evolutionary masses and/or underestimation of luminosities of the B supergiant sample. Data from spectral analysis by \cite{McEvoy} highlights that in most cases the spectroscopic masses of B supergiants are lower than the predicted evolutionary masses. This means in the M-L plane the data lies to the right when using spectroscopic masses, suggesting a more evolved stage than if the evolutionary masses had been utilised, given their systematically higher mass estimates and lying closer to the ZAMS in the M-L plane.

Finally, we compare the O supergiant evolutionary masses with the B supergiant spectroscopic masses in Fig. \ref{MspecBMevolO}, finding that the 2 populations now lie adjacent to one another, as would be expected in a HRD. This suggests that while the uncertainties in the O supergiant spectroscopic masses are large, the implications of B supergiant evolutionary masses may be key for studying the MS-width. For instance, inputs for internal mixing (\aov) directly impact the extension of the MS and the age inferred. The M-L plane allows \aov\ to be directly constrained to fit the observed  $M$, $L$, and $T_{\rm {eff}}$ giving a more robust age estimate. If a larger mixing efficiency is selected this may directly extend the MS to enclose the B supergiants. This means that their evolutionary masses should adopt a H-burning evolutionary stage rather than a post-MS assumption. 

We have compared the full sample of O and B supergiants in order to better understand the inconsistencies in spectroscopic and evolutionary masses of O and B supergiants, finding large errors in the spectroscopic masses of the O supergiant sample. We also find that evolutionary masses of B supergiants may not be appropriate since their evolutionary status is unknown and could lead to large discrepancies in the mass estimates. We therefore have used the M-L plane to test the effectiveness of each method of calculating stellar mass. Ultimately the key will be to use eclipsing binaries with dynamical masses to calibrate the internal mixing for precise ageing of stars.

\begin{figure}
\centering
\includegraphics[width = \columnwidth]{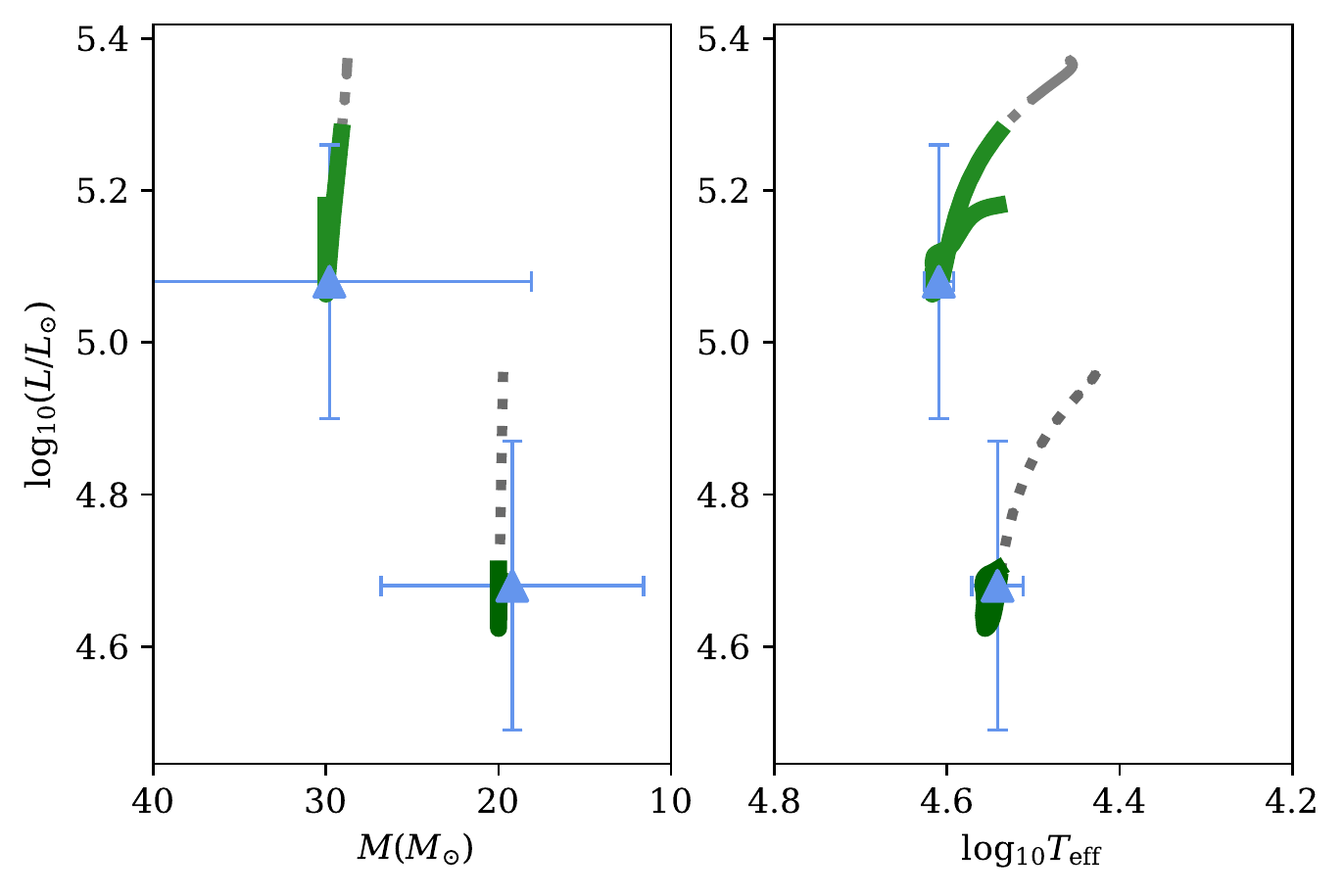}
\caption[VFTS 642]{\footnotesize Evolutionary tracks of VFTS 642 with initial masses of 30\Mdot\ and 20\Mdot, with initial rotation rates of $\Omega$/$\Omega_{\rm crit}$  $=$ 0.1 and \aov $=$ 0.1 in both cases. Green solid lines represent the evolutionary tracks until the observed effective temperature is reached, while grey dashed lines show the full MS evolution. Blue triangles represent the observed stellar parameters of VFTS 642 taken from \cite{TMBM3}.}
\label{642}
\end{figure}

\begin{figure}
\centering
\includegraphics[width = \columnwidth]{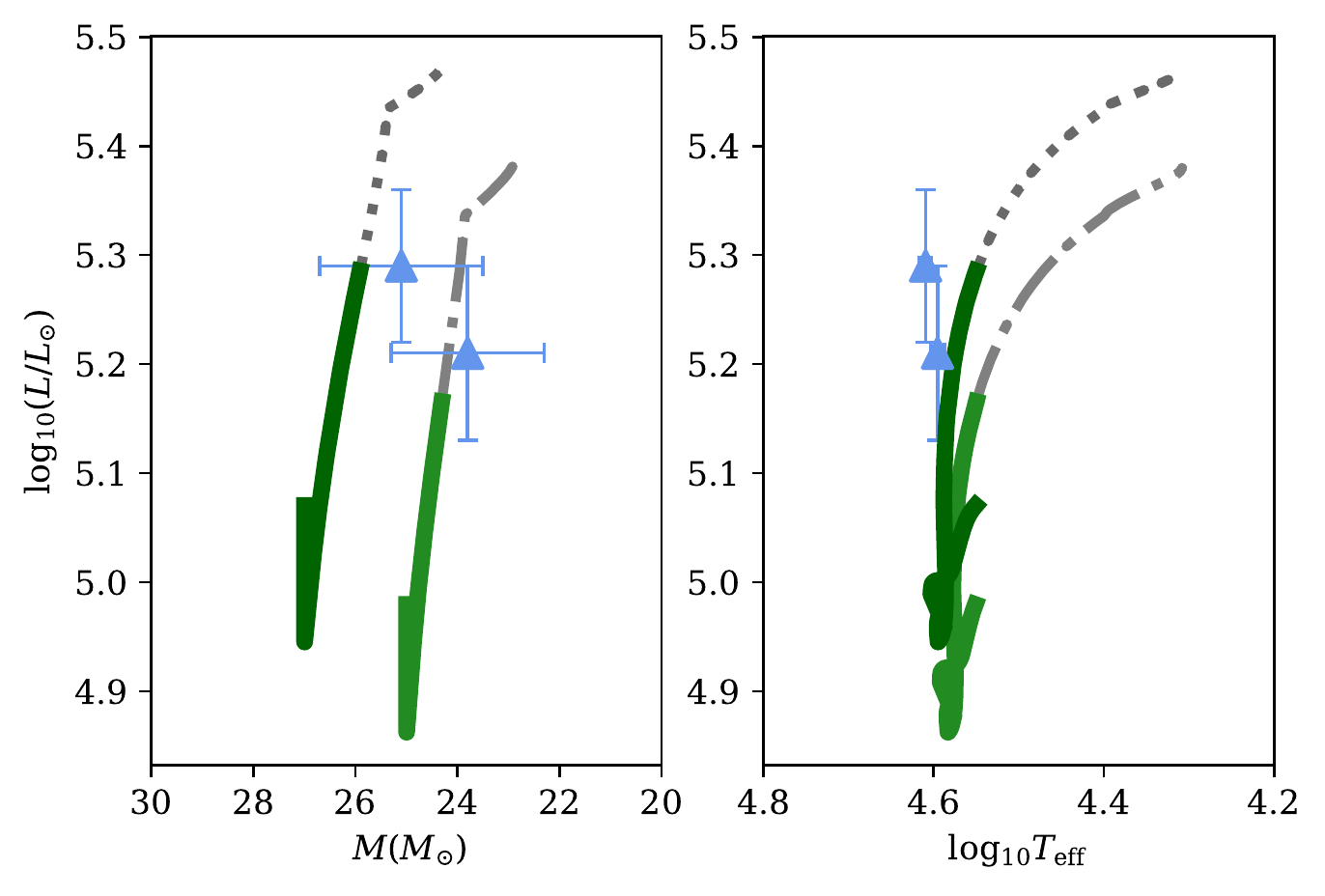}
\caption[VFTS 500]{\footnotesize Evolutionary models for VFTS 500 with initial masses of 27\Mdot\ and 25\Mdot\ for the primary and secondary respectively. Increased core overshooting is required of \aov $=$ 0.5 and initial rotation of $\Omega$/$\Omega_{\rm crit}$ $=$ 0.4 for both components is required. Green solid lines represent the evolutionary tracks until the observed effective temperature is reached, while grey dashed lines show the full MS evolution. Blue triangles represent the observed stellar parameters of VFTS 500 adapted from \cite{TMBM3}.}
\label{500}
\end{figure}
\subsection{Detached binary systems}\label{binaries}
We now provide examples of reliable M-L plane age constraints from detached eclipsing binaries. They are the most accurate due to their dynamical masses and since they have likely evolved from the same age. We provide 2 systems, 1 which lies close to the TAMS, and 1 which lies close to the ZAMS. These systems highlight that different amounts of mixing, via rotational mixing and overshooting, are required for different stars, possibly depending on their initial mass range or evolutionary stage (i.e. at the beginning of H-burning or near H-exhaustion). Moreover, after calibrating the M-L vector length, we can then compare our model's surface rotation rate after spin down due to stellar winds with the observed \vsini\ as a confirmation of both the mass-loss rates and the initial rotation rate. %We find that VFTS 527 showcases this very well. The 2 detached binary systems provide a wide range of masses (20-70\Mdot) while also probing the ZAMS and TAMS age estimates for these stars. 
Our new method indicates that estimating the amount of internal mixing is sensitive near the TAMS (\aov\ and $\Omega$/$\Omega_{\rm{crit}}$). However, we find that since stars spend a fraction of their MS lifetime at the ZAMS, it is difficult to constrain the amount of mixing close to the ZAMS as the vector length has not been well established.

We have analysed the evolution of VFTS 642 with the M-L plane, utilising dynamical masses, observed luminosities, and effective temperatures. We estimated the initial masses for the primary and secondary to be 30\Mdot\ and 20\Mdot\ respectively. These initial masses are evaluated based on the mass-luminosity relation which sets the ZAMS mass coupled with a steep $\dot{M}$ gradient in the M-L plane, due to Z-dependent winds at low $Z$ (50\% \Zdot). Figure \ref{642} illustrates the evolution of the components in both the M-L plane and HRD. We find that since the components are very close to the ZAMS, we required an initial rotation of $\Omega$/$\Omega_{\rm crit}$ = 0.1 and \aov\ $=$0.1, providing an estimated age of 2.3 Myr.  %Due to the effects of stellar winds, angular momentum is lost throughout the MS and by core H-exhaustion the surface rotation rate is approximately 60\kms in line with observations due to spin down with mass loss. 

%Since rotational mixing provides a significant extension to the evolutionary vector in the M-L plane, we test our evolutionary model with 40\% critical rotation with the observed effective temperature to evaluate whether additional mixing is required. When comparing our evolutionary track to the stellar observables, we found that minimal overshooting was required in both cases with \aov $=$ 0.1. Fig. \ref{527} suggests that both components of VFTS 527 are close to the TAMS, with our calibrated models predicting a current age of 2.9 $\pm$ 0.1 Myrs. 

Similarly, we provide estimates of the evolution for the detached binary VFTS 500, a detached eclipsing binary from the TMBM sample \citep{TMBM3}. We selected this system due to its proximity to the TAMS. Figure \ref{500} illustrates the evolution of VFTS 500 with our mixing-corrected evolutionary tracks. We find that the initial mass of the primary and secondary are 27\Mdot\ and 25\Mdot\ respectively. We estimate the initial rotation rates to be $\Omega$/$\Omega_{\rm crit}$ $=$ 0.4, with both components also requiring extra mixing via convective core overshooting equivalent to \aov $=$ 0.5, providing a current age of 6.4 Myr. Interestingly, we find that a larger extension by overshooting was necessary for reproducing VFTS 500 components compared with VFTS 642 components, suggesting that larger samples of detached eclipsing binaries with a variety of near-ZAMS and near-TAMS objects should be investigated. Currently, results from asteroseismology predict a range of internal mixing efficiencies by overshooting, see for example Table 1. from \cite{bowman20}.

% \textbf{While our analysis suggests 2 different determinations of \aov\ for VFTS 500 components and VFTS 642, we find that it is difficult to probe the full extent of \aov\ at the ZAMS since the full MS evolution has not yet been completed. Similarly, we find the eextent of interior mixing to be more sensitive at the TAMS, with these objects providing more reliable solutions. We therefore expect a slightly larger uncertainty on the \aov\ $=$ 0.1 finding for the components of VFTS 642. Moreover, studies by \cite{clarettorres16} and \cite{scott21} suggest that the amount of core overshooting may depend on evolutionary stage, and the difference in \aov\ values inferred from analysis of VFTS 500 and VFTS 642 may also be an artefact of stellar age itself, which should be tested in future works.}

\section{Summary}\label{conclusions}
We provide a summary of our results below:
\begin{itemize}
\item{We present a new tool to calculate stellar age in the M-L plane, tested in this work for single O stars, a large population of O and B supergiants, and finally with detached eclipsing binaries.}
\item{We show the inherent uncertainties in the isochrone-fitting method, mainly in adopting a standardised grid of models. We provide estimates for the key systematic and empirical uncertainties, confirming that the chosen \aov\ of the grid of models invokes significant errors on the inferred stellar age.}
\item{We demonstrate that spectroscopic masses and evolutionary masses are in disagreement in the M-L plane. We use the M-L plane to determine which method of determining stellar mass is most appropriate.}
\item{We find that the spectroscopic masses of the O supergiant sample are incorrect. We similarly find that the evolutionary masses of the B supergiant sample are inappropriate since their evolutionary stage is unknown.}
\item{We find that both components of VFTS 500 require high core overshooting (\aov\ $=$ 0.5), in agreement with \cite{Higgins}. As such, we explore an extended MS-width by including \aov\ $=$ 0.5, finding that most B supergiants are enclosed in the MS band, suggesting they could be core H-burning objects.}
\item{We reproduce the evolution of 2 detached eclipsing binaries, since these accurate dynamical mass measurements can provide precise age determinations in the M-L plane, while also providing robust internal mixing constraints.}
\item{We find that the M-L plane method of calculating stellar age as a function of internal mixing is most accurate near TAMS.}
\end{itemize}

In this work, we provide a new method of calculating the age of stars in the 'Mass-Luminosity' plane, as a function of observed luminosities, dynamical masses, and effective temperatures. We have utilised the M-L plane tool from \cite{Higgins} to now determine the age of stars, from a mixing-corrected model. The longstanding method of fitting isochrones of evolutionary model grids to observations can provide stellar age estimates, though with large systematic uncertainties. These errors are mainly due to the standardisation of stellar evolution grids which invoke a default mixing efficiency which has usually been calibrated for 1 mass, in most cases for low masses (1.5-20\Mdot). We provide comparisons of our updated ageing method to the previous isochrone method, finding errors of up to 3.5 Myr or 25\% of the MS lifetime. The consequences of such systematic uncertainties are far reaching, and in addition to empirical uncertainties, should be avoided by proper calibration of input physics in theoretical model grids. Particularly with the advancement of asteroseismology in massive star studies, we may begin to improve our model assumptions to better reflect the interior structure and evolution of stars as they would be in Nature, since we expect that stars will have different amounts of interior mixing based on their mass, age, metallicity and evolutionary stage.

We demonstrate our new ageing method in the M-L plane for O supergiants VFTS 109 and VFTS 306, with comparisons to isochronal ages, calculated for 2 assumptions of \aov\ (0.1 and 0.5), which naturally shorten or extend the MS lifetime. We provide detailed empirical and systematic errors on the stellar age from the analysis of VFTS 109, where the M-L plane age is 14.81 Myr, and the largest error is due to \aov\ giving an error of 3.1 Myr. The full method of correcting the internal mixing of each theoretical model to an observation in M-L space, where the observed luminosity and effective temperature are reached simultaneously with the mass, is provided for VFTS 306 where 4 extreme model tracks are compared for a range of \aov\ and rotation rates. We find an age of 5.64 Myr for VFTS 306 which is in reasonable agreement with the isochronal age estimates. This agreement is a result of having longer MS lifetimes for lower mass stars ($<$ 30\Mdot) leading to lower absolute errors (0.1-0.5 Myr) in stellar age, though still on the order of 10\% of the MS lifetime.

The M-L plane has been showcased in providing new estimates of stellar age, though this method inherits a drawback of using stellar masses to constrain evolutionary models to observed data, since the mass is not directly measured but inferred from spectral analysis or from comparisons to evolutionary models. The discrepancy noticed when comparing spectroscopic masses and evolutionary masses has led to a discord between theory and observations. We use the M-L plane to determine which method of estimating stellar mass may be more accurate and how we can infer which mass estimate is more reliable.

We compare the VFTS observations of O and B supergiants in the M-L plane, finding that the spectroscopic masses of the O supergiant sample from \cite{Ramirez13} lie in the forbidden region of the M-L plane, below the ZAMS determined by the mass-luminosity relation, suggesting that the O supergiant spectroscopic masses are incorrect. We then compare the evolutionary masses of the O and B supergiant sample finding that the samples now overlay as 1 population. While investigating the position of the O and B stars in the M-L plane, we discovered a lack of distinction between the spectral types, as a result of employing evolutionary masses taken from \cite{McEvoy}. We expect that this error is due to invoking an evolutionary status on B supergiants which largely have unknown evolutionary stages (core H-burning or core He-burning). Finally, when comparing with spectroscopic masses of the B supergiant sample and the evolutionary masses of the O supergiant sample in the M-L plane, we found that the B supergiants now lie beyond the O star range, as would be expected in a HRD, suggesting that most B supergiants may still be H-burning objects. In fact, when comparing the O and B supergiant samples in a standard HRD, implementing \aov $=$ 0.5 which extends the MS-width, we find that over 80\% of the B supergiant sample are now enclosed (except for 6 stars) in the MS-width. From \cite{Higgins}, we find this value of \aov\ to be appropriate in the 30-40\Mdot\ range in order to reproduce the evolution of the detached eclipsing binary HD166734. So while it remains unclear which \aov\ is appropriate in various mass ranges, there is a possibility that B supergiants are still core H-burning objects.

In order to precisely measure the stellar mass and avoid the discrepancy outlined above, we rely on dynamical masses from detached eclipsing binaries. Such accurate masses may provide precise age estimates in the M-L plane, where the internal mixing efficiencies can also be constrained for similar mass ranges and evolutionary stages \citep{south22}. We therefore reproduce the evolution of 2 detached systems from the TMBM sample, VFTS 642 and VFTS 500. We select these binary systems to provide a range of current masses, and to probe the effectiveness of our method at near-ZAMS and near-TAMS locations. Having dynamical masses from \cite{TMBM3}, we can utilise these binary systems as in \cite{Higgins} to test stellar evolution now in the LMC. We find that in order to reproduce the mass and luminosities of VFTS 642 at the vector length of the observed temperatures, \aov\ = 0.1 is required, having initially constrained the rotation rates to $\Omega$/$\Omega_{\rm crit}$ = 0.1, giving an age of 2.3 Myr with initial masses of 30\Mdot\ and 20\Mdot\ for the primary and secondary component respectively. Similarly, we reproduce the evolution of VFTS 500 finding initial masses of 27\Mdot\ and 25\Mdot\ for the primary and secondary, and a current age of 6.4 Myr. We estimate initial rotation rates of 40\% critical rotation for the primary and secondary, with both components requiring \aov\ $=$ 0.5. Interestingly, we find that our M-L plane method is most useful near the TAMS since the vector length can be probed more accurately as a function of the MS lifetime and interior mixing, whereas close to the ZAMS the full vector length is not yet realised and is more challenging for determining the precise mixing efficiencies.

In summary, we provide multiple test-cases for our new method of determining the age of individual stars, with mixing-corrected models. For future observations of massive stars with ULLYSES, XShooter and WEAVE, accurate mass and age determinations will be important, particularly in  resolving the mass discrepancy problem and the B supergiant problem. We know that assuming a constant internal mixing efficiency for all masses will impose systematic uncertainties on the age of stars, particularly lower mass stars ($<$ 30\Mdot) which are significantly more numerous, therefore we must attempt a better fit of our theoretical models.

\section*{Acknowledgements}
The authors acknowledge MESA authors and developers for their continued revisions and public accessibility of the code. JSV and ERH are supported by STFC funding under grant number ST/V000233/1 in the context of the BRIDGCE UK Network. EH would like to thank Gavin Ramsay and Jose Groh for constructive discussions which improved the development of the manuscript.
%%%%%%%%%%%%%%%%%%%%%%%%%%%%%%%%%%%%%%%%%%%%%%%%%%
\section*{Data Availability}

The data underlying this article will be shared on reasonable request
to the corresponding author.
\typeout{}

\bibliographystyle{mnras}
\bibliography{newdiff.bib}

% \begin{appendix}\label{gridapp}

% \end{appendix}

\end{document}